\documentclass[%
 aip,
 amsmath,amssymb,
 reprint,%
]{revtex4-1}

\usepackage{graphicx}
\usepackage{dcolumn}
\usepackage{bm}

\usepackage[utf8]{inputenc}
\usepackage[T1]{fontenc}
\usepackage{mathptmx}
\usepackage{amsthm}
\usepackage{amsmath,amssymb,amsfonts,mathtools}
\usepackage{etoolbox}
\usepackage{physics}
\usepackage[utf8]{inputenc}
\usepackage{tikz}
\usetikzlibrary{positioning}
\usetikzlibrary{calc}
\newtheorem{definition}{Definition}
\newtheorem{remark}{Remark}
\newtheorem{proposition}{Proposition}
\makeatletter
\def\@email#1#2{%
 \endgroup
 \patchcmd{\titleblock@produce}
  {\frontmatter@RRAPformat}
  {\frontmatter@RRAPformat{\produce@RRAP{*#1\href{mailto:#2}{#2}}}\frontmatter@RRAPformat}
  {}{}
}%

\usepackage[normalem]{ulem}

\makeatother
\begin{document}

\preprint{AIP/123-QED}

\title[Fractional dynamics]{A Fractional Calculus Framework for Open Quantum Dynamics: From Liouville to Lindblad to Memory Kernels}
\author{Bo Peng}
 \email{peng398@pnnl.gov}
\affiliation{Physical and Computational Sciences Directorate, Pacific Northwest National Laboratory, Richland, WA, 99354, USA.
}%

\author{Yu Zhang}
\affiliation{%
Theoretical Division, Los Alamos National Laboratory, Los Alamos, NM, 87544, USA.
}%

\date{\today}

\begin{abstract}
Open quantum systems exhibit dynamics ranging from unitary evolution to irreversible dissipation. While the Gorini--Kossakowski--Sudarshan--Lindblad (GKSL) equation uniquely characterizes Markovian CPTP evolution, many physical platforms display non-Markovian features such as algebraic relaxation and coherence backflow. Fractional calculus provides a natural way to model such long-memory behavior through power-law temporal kernels introduced by fractional time derivatives. Here we develop a unified framework that embeds fractional master equations within the broader hierarchy of open-system formalisms. The fractional equation forms a structured subclass of memory-kernel models, reduces to the Lindblad form at unit order, and, through Bochner--Phillips subordination, admits a CPTP representation as an average over Lindblad semigroups. \textcolor{black}{Its resolvent structure further connects fractional dynamics to established non-Markovian approaches, including Nakajima--Zwanzig kernels and hierarchical equations of motion, providing a compact surrogate for long-memory effects.} This formulation positions fractional calculus as a rigorous and practical language for quantum dynamics with intrinsic memory, supporting both analytical insight and efficient quantum simulation.
\end{abstract}

\maketitle

\section{Introduction}

The dynamics of an isolated quantum system are governed by the Liouville--von Neumann equation, which generates perfectly unitary and reversible evolution~\cite{vonNeumann1929}. When a system interacts with its environment, however, the reduced dynamics generally become irreversible and often highly nontrivial. The celebrated Gorini--Kossakowski--Sudarshan--Lindblad (GKSL) equation~\cite{Gorini1976,Lindblad1976} provides the unique characterization of \emph{Markovian} open system dynamics, ensuring CPTP semigroup evolution~\cite{Alicki1977}. Yet, many experimental platforms--from superconducting qubits~\cite{Murch2013,Kjaergaard2020}, nitrogen-vacancy centers~\cite{PhysRevLett.121.060401}, trapped ions~\cite{Clos2016}, and ultracold gases~\cite{PhysRevA.87.012127}---display non-Markovian behavior such as non-exponential decay, coherence backflow, and power-law tails, indicating the presence of memory effects beyond the reach of standard Lindblad dynamics~\cite{Vacchini2011,Dominy2015,Chruscinski2017}.

Fractional calculus has emerged as a powerful framework for modeling such long-memory and anomalous relaxation phenomena~\cite{Hilfer2000,Tarasov2021}. By replacing the first-order derivative with a fractional time derivative, one introduces a power-law memory kernel that endows the dynamics with nonlocal temporal correlations. The corresponding decay, governed by the Mittag--Leffler function~\cite{Mainardi2000}, interpolates continuously between exponential and algebraic regimes. Despite its empirical success in describing slow relaxation and heavy-tailed dynamics, most fractional formulations in quantum contexts remain phenomenological, lacking a rigorous connection to established open system theory~\cite{Breuer2002}. Fundamental questions persist regarding how fractional evolution relates to the GKSL semigroup, whether it preserves complete positivity, and how it fits within the general class of memory-kernel quantum master equations (QMEs).

\textcolor{black}{Addressing these questions requires placing fractional dynamics within the broader QME landscape. Although numerically exact open-system solvers have advanced rapidly, QMEs themselves remain central because they provide analytical insight, scalable reduced descriptions, and generator-level models that interface naturally with quantum simulation and control frameworks. Clarifying where fractional dynamics sit within this landscape therefore involves relating them to existing QME-based approaches and understanding how they complement, rather than compete with, exact computational methods.}

\textcolor{black}{
A wide variety of non-Markovian generalizations of the GKSL equation have been proposed. Projection-operator methods such as the Nakajima--Zwanzig (NZ) equation~\cite{Nakajima1958,Zwanzig1960} introduce memory kernels through explicit system--bath couplings, while the time-convolutionless (TCL) formulation yields time-dependent generators that may lose CP-divisibility~\cite{Breuer2002,Breuer2016}. Generalized Lindblad approaches (e.g., Ref.~\citenum{Breuer2007GeneralizedLindblad}) incorporate environmental correlations via correlated projection operators or extended state spaces, and typically retain a first-order time derivative but with time-dependent decay rates. Microscopic approaches such as hierarchical equations of motion (HEOM)~\cite{HEOM_1,HEOM_2,HEOM_3,HEOM_4} provide numerically exact descriptions for Gaussian environments but at a high computational cost. These formalisms collectively establish a rich landscape of non-Markovian models,
and recent years have seen substantial progress in improving their numerical efficiency, for example through tensor-network representations~\cite{Chen:2025to, Yan:2021tg}. Nevertheless, even in these advanced implementations, they generally lack the simplicity of a closed-form CPTP propagator or an analytically transparent interpolation between Markovian and memory-dominated regimes.}

\textcolor{black}{Beyond projection-operator and time-local master equations, many non-Markovian methods arise from influence-functional formalisms, including the quasi-adiabatic propagator path integral (QUAPI)~\cite{Makri1995QUAPI1, Makri1995QUAPI2}, multiconfiguration time-dependent Hartree (MCTDH) approaches to open-system dynamics~\cite{Thoss2001MCTDH, Wang2003MCTDHReview}, and related path-integral techniques. These frameworks provide controlled microscopic descriptions of dissipative dynamics but typically involve large tensor-network objects or exponential memory growth as temporal correlations accumulate. In contrast, the fractional approach developed here provides a compact, closed-form surrogate for environments with slow or algebraically decaying correlations. While it does not reproduce the full influence-functional structure, it captures the emergent memory behavior in a generator-level deformation of the Lindblad semigroup. Thus, fractional dynamics offer an analytically transparent complement to computationally intensive influence-functional methods, providing a reduced model that summarizes the dominant memory features encoded in more detailed path-integral treatments.}

\textcolor{black}{The above considerations motivate the development of a framework that situates fractional evolution within the established hierarchy of open-system formalisms while retaining the interpretability of generator-based descriptions.} 
In this work, we develop a unified theoretical framework that places fractional dynamics within the broader hierarchy of quantum open system formalisms. We show that fractional master equations form a structured subclass of non-Markovian QMEs, coinciding with standard Lindblad dynamics in the limit of unit fractional order. This perspective reveals that fractional differentiation introduces memory at the level of the generator rather than through explicit environmental coupling, bridging the conceptual gap between purely unitary and dissipative descriptions.

\textcolor{black}{
Our key observation is that the fractional master equation admits a resolvent-level representation $(s^\alpha\mathbb{I}-\mathcal{L})^{-1}$ that can be interpreted through Bochner--Phillips subordination~\cite{Bochner1955,Feller1971} as a convex mixture of Lindblad semigroups evaluated at L\'{e}vy-distributed operational times. This structure guarantees complete positivity for any GKSL generator and provides a closed analytical solution in terms of the Mittag--Leffler function. Moreover, we show that the fractional resolvent is algebraically connected to the Laplace-resolvent forms appearing in the NZ equation (long-tailed kernels), HEOM (self-energy corrections), and time-local generalized Lindblad approaches (time-dependent rates). In this sense, fractional dynamics emerge not as a phenomenological alteration of quantum mechanics but as a mathematically precise deformation of the Lindblad semigroup that captures algebraic memory in a compact, two-parameter form.}

Together, these insights establish a coherent storyline that connects the principal regimes of quantum dynamics: unitary Liouville evolution forms the base, Lindbladian semigroups represent Markovian open systems, and fractional extensions introduce structured non-Markovianity through power-law memory kernels. This unified framework provides both conceptual clarity and a mathematical foundation for simulating long-memory quantum processes on emerging quantum and hybrid computing platforms. Unlike phenomenological fractional models, the present formulation embeds fractional differentiation directly within the GKSL generator and preserves complete positivity through Bochner–Phillips subordination~\cite{Bochner1955,Feller1971}.
This distinguishes it from projection-operator~\cite{Nakajima1958,Zwanzig1960} and time-convolutionless approaches~\cite{Breuer2002,Breuer2016}, which introduce memory through explicit environmental couplings rather than at the generator level.


\section{Theory}

\subsection{Preliminaries}

Let $\mathcal{H}$ be a separable Hilbert space, and let $\mathcal{B}_1(\mathcal{H})$ denote the Banach space of trace-class operators acting on $\mathcal{H}$.  A physical quantum state is a density operator $\rho\in\mathcal{B}_1(\mathcal{H})$ satisfying $\rho\ge0$ and $\mathrm{Tr}(\rho)=1$.  The time evolution of $\rho$ is represented by a family of linear maps $\{\Phi(t)\}_{t\ge0}$ acting on $\mathcal{B}_1(\mathcal{H})$.  Physical consistency requires that each $\Phi(t)$ be CPTP; this condition ensures that the evolution of $\rho$ remains valid even when the system is part of a larger entangled state, and is thus central to any legitimate description of quantum dynamics~\cite{Nielsen2000,Chruscinski2017}.

When the system is isolated, the evolution is unitary and memoryless, governed by the Liouville--von Neumann equation~\cite{vonNeumann1929}.  The corresponding maps form a one-parameter unitary group that is CPTP, invertible, and norm-preserving.  In the presence of environmental coupling, irreversibility and decoherence arise.  If the dynamics remain time-homogeneous and memoryless, the family $\{\Phi(t)\}$ constitutes a \emph{quantum dynamical semigroup}~\cite{Alicki1977}:
\begin{align}
    \Phi(0)=\mathbb{I},\qquad \Phi(t+s)=\Phi(t)\Phi(s), \label{eq:composition}
\end{align}
with $\Phi(t)$ CPTP and norm-continuous.  The Gorini--Kossakowski--Sudarshan--Lindblad (GKSL) theorem guarantees that any such semigroup admits a generator $\mathcal{L}$ of Lindblad form~\cite{Gorini1976,Lindblad1976}, uniquely characterizing \emph{Markovian} open system dynamics in which all memory of past states is lost.

In many physical systems, this Markovian approximation breaks down.  Coherence revivals, stretched-exponential relaxation, and algebraic long-time tails all indicate that the generator $\mathcal{L}$ becomes effectively time-nonlocal~\cite{Breuer2016,Li2018,deVega2017}.  The resulting dynamics are better described by \emph{memory-kernel} QMEs, which introduce explicit temporal correlations through convolution integrals~\cite{Nakajima1958,Zwanzig1960}.  Fractional calculus provides a compact and rigorous formulation of such dynamics: fractional time derivatives generate power-law kernels that naturally give rise to Mittag--Leffler relaxation, interpolating between exponential and algebraic behavior~\cite{Mainardi2000,Gorenflo2014,Luchko2020}.  In the limit $\alpha~\to~1$, the fractional derivative recovers the Markovian Lindblad semigroup; for $\alpha<1$, the dynamics become non-divisible yet remain CPTP and physically consistent~\cite{Chruscinski2017,Hall2014}.

Fractional dynamics therefore provide a unified mathematical bridge connecting distinct regimes of quantum evolution—from unitary to non-unitary, from memoryless to memoryful—under a single operator-theoretic framework.  To make this connection precise, we introduce the following hierarchy of dynamical models.

\begin{definition}[Liouville Dynamics]
\label{def:Liouville}
\emph{Liouville dynamics} describe the unitary evolution of a closed quantum system governed by a Hermitian Hamiltonian $H=H^\dagger$:
\begin{align}
    \dot{\rho}(t) = -i[H,\rho(t)].
\end{align}
The maps $\Phi(t)(\rho)=e^{-iHt}\rho e^{iHt}$ form a one-parameter group of CPTP and norm-preserving transformations.
\end{definition}

\begin{definition}[Lindblad/GKSL Dynamics]
\label{def:Lindblad}
\emph{Lindblad dynamics} generalize Liouville dynamics to Markovian open systems.  
They satisfy
\begin{align}
    \dot{\rho}(t)=\mathcal{L}\rho(t), \label{eq:Lindblad1}
\end{align}
with
\begin{align}
    \mathcal{L}(\rho)=-i[H,\rho]+\sum_j\gamma_j~\left(L_j\rho L_j^\dagger-\tfrac12\{L_j^\dagger L_j,\rho\}\right). \label{eq:Lindblad2}
\end{align}
Here, $\gamma_j\ge0$ quantifies the strength of the dissipative channel represented by the jump operator $L_j$.  
The solution $\rho(t)=e^{t\mathcal{L}}\rho(0)$ defines a CPTP semigroup, and complete positivity is ensured precisely by the Lindblad structure of $\mathcal{L}$ superoperator.
\end{definition}

\begin{definition}[Memory-Kernel Quantum Master Equation]
\label{def:QME}
A \emph{convolution-type} or \emph{memory-kernel} quantum master equation describes non-Markovian dynamics through an integral kernel $K(t-\tau)$:
\begin{align}
\dot{\rho}(t)=\int_0^t K(t-\tau)~\mathcal{L}\rho(\tau)~d\tau.
\end{align}
The kernel encodes temporal correlations and environmental backaction.  
Choosing $K(t)=\delta(t)$ recovers Lindblad dynamics, while more general kernels yield stretched-exponential or power-law relaxation~\cite{Breuer2002,deVega2017}.
\end{definition}

\begin{definition}[Fractional Master Equation]
\label{def:Fractional}
For $0<\alpha<1$, a \emph{fractional master equation} replaces the first-order time derivative with a fractional (Caputo) derivative of order $\alpha$:
\begin{align}
\prescript{C}{}{D}_t^{\alpha}\rho(t)=\mathcal{L}\rho(t). \label{eq:FME}
\end{align}
Here $\mathcal{L}$ is a densely defined linear generator on $\mathcal{B}_1(\mathcal{H})$.  
The fractional derivative introduces a long-range power-law memory kernel, giving rise to Mittag--Leffler relaxation that interpolates between exponential and algebraic decay.  
In the limit $\alpha~\to~1$, the equation reduces to the standard first-order master equation $\dot{\rho}(t)=\mathcal{L}\rho(t)$~\cite{Podlubny1999,Tarasov2021}.
\end{definition}

\begin{remark}[\textcolor{black}{\textbf{Physical meaning of the fractional order \(\alpha\).}}]
\leavevmode

\noindent
\textcolor{black}{The fractional order \(0<\alpha<1\) has a clear operational interpretation within the renewal–process framework underlying Eq.~\eqref{eq:FME}. In Bochner--Phillips subordination, the physical time \(t\) is replaced by a random operational time \(U(t)\) associated with an inverse–stable subordinator~\cite{Bochner1955,Feller1971,Schilling2012Bernstein,BaeumerMeerschaert2001}. The corresponding renewal process has a heavy–tailed waiting–time density \(\psi_\alpha(\tau)\sim \tau^{-1-\alpha}\), a structural consequence of the L\'{e}vy measure of the underlying \(\alpha\)-stable subordinator~\cite{METZLER20001,MainardiScalas2004,MeerschaertSikorskii2012}. The parameter \(\alpha\) therefore controls the heaviness of the waiting-time tail and thus the strength and persistence of memory in the stochastic time change.  When \(\alpha=1\), the waiting times become exponential and one recovers the Markovian Lindblad semigroup; when \(\alpha<1\), the broad, power-law waiting statistics generate non-Markovian dynamics with Mittag--Leffler relaxation, stretched-exponential transients, and long-time algebraic tails.}

\textcolor{black}{In this sense, \(\alpha\) is \emph{not a phenomenological constant} but a structural memory exponent: it determines how far the dynamics deviate from Markovian behavior and how quickly the influence of past states decays.  In microscopic settings such as the pure-dephasing spin--boson model, the value of \(\alpha\) can be related to properties of the bath correlation function.  Sub-Ohmic baths, which yield slow algebraic decay of coherence, correspond to smaller \(\alpha\), while Ohmic or super-Ohmic baths, which produce faster decay, yield \(\alpha\) closer to unity.  Thus, \(\alpha\) provides a compact and operational measure of bath-induced memory strength that connects the fractional master equation directly to physically observable decoherence profiles.}
\end{remark}

\begin{remark}[Fractional Liouville and Fractional Lindblad]
\leavevmode

\noindent
(i) \emph{Fractional Liouville dynamics} correspond to the closed-system case where $\mathcal{L}=-i[H,~\cdot~]$ with $H=H^\dagger$. They describe coherent yet memoryful evolution that remains CPTP but non-divisible, effectively a continuous mixture of unitaries~\cite{Bochner1955,Feller1971}.

\noindent
(ii) \emph{Fractional Lindblad dynamics} arise when $\mathcal{L}$ takes the GKSL form.  
These dynamics represent physically consistent non-Markovian extensions of standard Lindbladian evolution, equivalent to subordinated Lindblad semigroups governed by power-law waiting-time distributions~\cite{Hall2014}.
\end{remark}

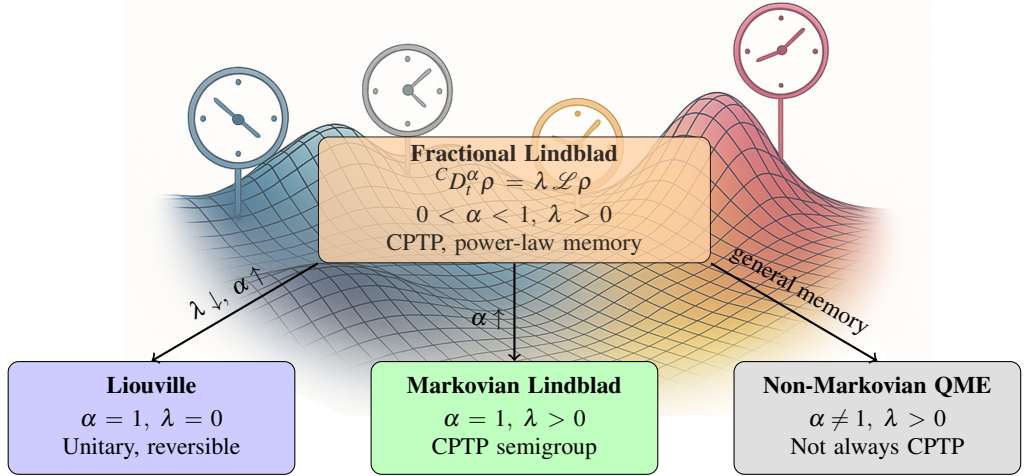
\begin{figure*}[t]
\centering
\begin{tikzpicture}[
    box/.style={
        draw,
        rectangle,
        minimum width=5cm,
        minimum height=2.5cm,
        text width=5cm,
        align=center,
        font=\sffamily,
        thick
    },
    arr/.style={
        {Stealth[length=3mm]}-{Stealth[length=3mm]},
        line width=1.5pt
    },
    node distance=1.5cm 
]

\tikzstyle{block}=[rectangle, rounded corners, draw=black, align=center, minimum width=3.8cm, minimum height=1.5cm, text width=3.6cm]
\tikzstyle{arrow}=[->, thick]

\node[opacity=0.8] at (5,3)
 {\includegraphics[width=0.6\linewidth,trim=0cm 5.5cm 0cm 3cm, clip]{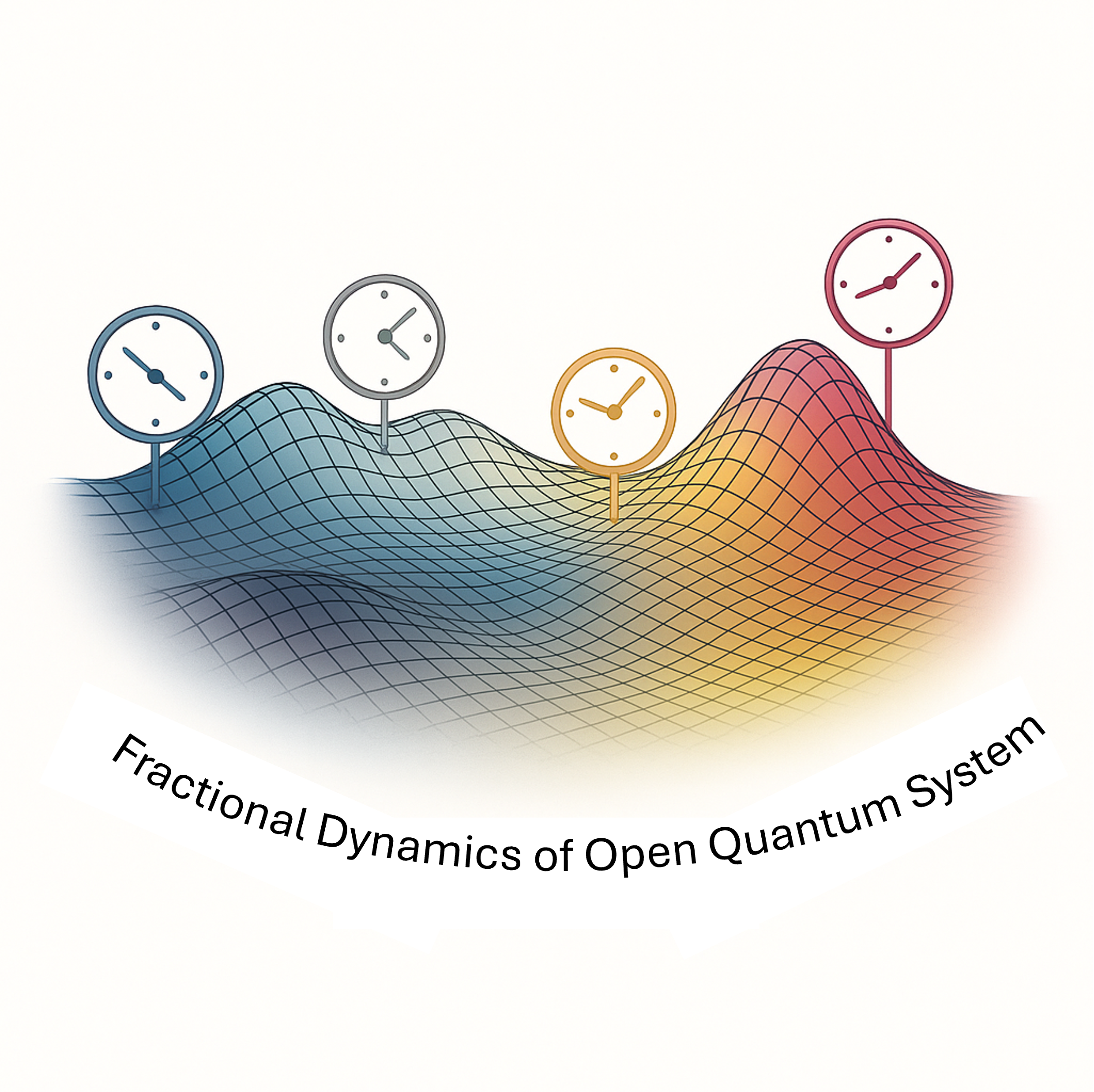}};

\node[block, fill=blue!20] (liouville) {\textbf{Liouville}\\[2pt]
$\alpha=1,\ \lambda=0$\\
Unitary, reversible};

\node[block, fill=green!25, right=of liouville, xshift=-0.5cm] (lindblad) {\textbf{Markovian Lindblad}\\[2pt]
$\alpha=1,\ \lambda>0$\\
CPTP semigroup};

\node[block, fill=gray!25, right=of lindblad, xshift=-0.5cm] (memory) {\textbf{Non-Markovian QME}\\[2pt]
$\alpha\!\neq\!1,\ \lambda>0$\\
Not always CPTP};

\node[opacity=0.8, block, fill=orange!35, above=of lindblad, yshift=-0.2cm, text width=5cm] (fractional) {\textbf{Fractional Lindblad}\\
$\prescript{C}{}{D}_t^{\alpha}\rho=\lambda\,\mathcal{L}\rho$\\[2pt]
$0<\alpha<1,\ \lambda>0$\\
CPTP, power-law memory};

\draw[arrow] (fractional.south west) -- node[above left, xshift=20pt, yshift=0pt, sloped]{$\lambda\downarrow,~\alpha\uparrow$} (liouville.north);
\draw[arrow] (fractional.south) -- node[above right, xshift=-20pt, yshift=-10pt]{$\alpha\uparrow$} (lindblad.north);
\draw[arrow] (fractional.south east) -- node[right, xshift=-35pt, yshift=8pt, sloped]{general memory} (memory.north);
 
\end{tikzpicture}

\caption{Schematic hierarchy of quantum dynamical regimes unified by fractional subordination. \textcolor{black}{The parameter $\alpha$ is the \emph{memory exponent} of the renewal process governing the operational time of the dynamics:} $\alpha=1$ yields Markovian Lindblad semigroups with exponential waiting times, while $0<\alpha<1$ generates long-tailed waiting statistics and non-Markovian evolution with algebraically decaying memory.  \textcolor{black}{The parameter $\lambda$ controls the dissipative strength of the underlying GKSL generator.}  Fractional or subordinated Lindblad dynamics ($0<\alpha<1$) therefore provide a CPTP bridge connecting unitary Liouville evolution ($\alpha=1,\lambda=0$), Markovian semigroups ($\alpha=1,\lambda>0$), and general memory--kernel quantum master equations ($\alpha\neq 1$).}
\label{fig:hierarchy}
\end{figure*}

These nested definitions establish the structural hierarchy as schematically illustrated in Figure~\ref{fig:hierarchy}, in which fractional calculus connects the memoryless semigroup of Lindblad evolution to the nonlocal, algebraic relaxation characteristic of non-Markovian quantum systems~\cite{Breuer2016,Li2018,deVega2017}.

The structural hierarchy above admits a natural mathematical interpretation through the concept of \emph{subordination}.  
In this picture, fractional dynamics arise as averages over standard semigroup evolutions with respect to a probability density in operational time~\cite{Bochner1955,Feller1971}.  
Specifically, the fractional flow can be represented as a Bochner--Phillips subordination of a Lindblad semigroup, in which the physical time $t$ is replaced by a random operational time distributed according to a power-law waiting-time function~\cite{Bochner1955,Feller1971,Mainardi2000}.  
This representation guarantees complete positivity whenever the underlying generator $\mathcal{L}$ is of GKSL form and provides a direct link between fractional differentiation, non-divisibility, and long-memory behavior~\cite{Rivas2014}.  
In the next section, we make this correspondence explicit and examine how fractional subordination generates algebraic relaxation and interpolates smoothly between unitary, Markovian, and non-Markovian regimes.

Fractional extensions of quantum master equations have been explored in various contexts, from phenomenological models of quantum dissipation with power-law memory~\cite{Tarasov2021} to more recent formulations of time-fractional Lindblad dynamics for weakly coupled systems~\cite{Ang2024MemoryInducedWeakDissipation}. Parallel developments have applied subordination principles in quantum settings, including Bochner and topological subordination for fractional evolution equations \cite{Feller1971,Sandev2018FCAA}. These approaches, while illuminating, often treat fractional derivatives or memory kernels as postulated modifications rather than consequences of a well-defined stochastic process.
In contrast, the present work provides a rigorous unification based on \emph{Bochner–Phillips subordination of Lindblad semigroups}. This construction ensures that the fractional propagator CPTP for any \(0<\alpha\le1\), thereby reconciling fractional dynamics with quantum dynamical semigroup theory. The probabilistic interpretation in terms of random operational times \(u \sim f_\alpha(u,t)\) further connects power-law relaxation directly to waiting-time statistics, clarifying the origin of algebraic tails in open system coherence.


\subsection{Fractional Master Equation}

Fractional calculus provides a rigorous way to encode long-range memory in quantum dynamics. In the \emph{fractional master equation}~\eqref{eq:FME}, $\mathcal{L}$ is a densely defined generator on $\mathcal{B}_1(\mathcal{H})$, and the Caputo derivative acts on differentiable functions $f:[0,\infty)\to \mathcal{B}_1(\mathcal{H})$ as
\begin{align}
\prescript{C}{}{D}_t^\alpha f(t)
&= \Big(I^{1-\alpha} \circ \frac{d}{dt}\Big) f(t) \notag \\
&= \frac{1}{\Gamma(1-\alpha)}\int_0^t \frac{f'(\tau)}{(t-\tau)^\alpha}~d\tau,
\qquad 0<\alpha <1, \nonumber\\
&\equiv \int \kappa_C(t-\tau) f'(\tau) d\tau,
\end{align}
where $\kappa_C(t) = \frac{t^{-\alpha}}{\Gamma(1-\alpha)}$ and $I^{1-\alpha}$ denotes fractional integration~\cite{Podlubny1999,Kilbas2006,Luchko2020}. The Gamma function
\begin{align}
\Gamma(z) = \int_0^\infty t^{z-1}e^{-t}~dt,
\qquad \Re(z)>0,
\end{align}
fixes the normalization of the weakly singular kernel $(t-\tau)^{-\alpha}$ in accordance with the Riemann–Liouville (RL) conventions~\cite{Kilbas2006}. In this way, an explicit power-law weighting of past rates $f'(\tau)$ replaces instantaneous evolution by a history-dependent response.

\subsubsection{Equivalence to convolution master equations}
\label{subsec:equiv-convolution}

Starting from Eq.~\eqref{eq:FME}, applying the RL integral $I_t^\alpha$ and using $\prescript{C}{}{D}_t^\alpha I_t^\alpha=\mathbb{I}$ gives the Volterra form
\begin{align}
\label{eq:volterra-main}
\rho(t)=\rho(0)+\int_0^t K_\alpha^{(\mathrm{V})}(t-\tau)~\mathcal{L}\rho(\tau)~d\tau
\end{align}
with the kernel
\begin{align}
K_\alpha^{(\mathrm{V})}(t)=\frac{t^{\alpha-1}}{\Gamma(\alpha)}.
\end{align}
Differentiating Eq.~\eqref{eq:volterra-main} (Leibniz rule for weakly singular kernels) yields the differential convolution form
\begin{align}
\label{eq:conv-main}
\dot{\rho}(t)=\int_0^t k_\alpha(t-\tau)~\mathcal{L}\rho(\tau)~d\tau
+\frac{t^{\alpha-1}}{\Gamma(\alpha)}~\mathcal{L}\rho(0)
\end{align}
with the kernel
\begin{align}
k_\alpha(t)=\frac{d}{dt}K_\alpha^{(\mathrm{V})}(t)=\frac{t^{\alpha-2}}{\Gamma(\alpha-1)}.\label{eq:Frac_kernel}
\end{align}
%
\textcolor{black}{
The boundary term in Eq.,\eqref{eq:conv-main}, $\frac{t^{\alpha-1}}{\Gamma(\alpha)}~\mathcal{L}\rho(0)$ arises generically when the Caputo form is converted into a differential convolution equation~\cite{Chruscinski2017,Tarasov2021}. For the fractional regime of primary interest in this work, $0<\alpha<1$, this term represents an \emph{initial-slip correction} reflecting the mismatch between the initial state and the stationary state of the Lindbladian. Its behavior is fully analogous to short-time slip terms in NZ memory-kernel master equations and in CTRW models of anomalous relaxation~\cite{Metzler2014}. Although $t^{\alpha-1}$ diverges as $t\rightarrow 0$ for $0<\alpha<1$, its integrated influence is finite and it decays rapidly, contributing only transiently before the Mittag–Leffler relaxation dominates. By contrast, in the short-time Zeno regime analyzed later the effective order turns to be $\alpha=2$; in this case the same prefactor scales as $t^{\alpha-1} = t$ and therefore \emph{vanishes} as $t\rightarrow 0$, so no divergence or anomalous enhancement occurs at early times.}

It is helpful to separate the roles of the three kernels that appear across these equivalent forms. The Volterra kernel $K_\alpha^{(\mathrm{V})}(t)=t^{\alpha-1}/\Gamma(\alpha)$ multiplies $\mathcal{L}\rho$ in the integral equation Eq.~\eqref{eq:volterra-main} and is completely monotone on $\mathbb{R}_+$; this complete monotonicity means $K_\alpha^{(\mathrm{V})}$ is the Laplace transform of a positive measure and underlies the subordination representation developed later. Differentiating $K_\alpha^{(\mathrm{V})}$ produces $k_\alpha(t)=t^{\alpha-2}/\Gamma(\alpha-1)$, the kernel in the differential convolution form Eq.~\eqref{eq:conv-main}. By contrast, the Caputo derivative itself can be written as a convolution of $\dot{\rho}$ with $\kappa_C(t)=t^{-\alpha}/\Gamma(1-\alpha)$; this kernel lives \emph{inside} the Caputo operator and should not be identified with the memory kernels that multiply $\mathcal{L}\rho$. Detailed proofs are provided in Appendix~\ref{app:caputo-volterra} (Proposition~\ref{prop:equiv}).

\subsubsection{Mittag--Leffler solution and physical interpretation}
The formal solution of the fractional master equation can be written using the operator Mittag--Leffler function as
\begin{align}
\rho(t)=E_\alpha(t^\alpha \mathcal{L})\,\rho(0),
\label{eq:Mittag-Leffler}
\end{align}
where the Mittag-Leffler function $E_\alpha(z)=\sum_{n=0}^{\infty} z^n/\Gamma(1+\alpha n)$ is defined via its convergent power series~\cite{Mainardi2000}. 
If $\mathcal{L}$ is diagonalizable with eigenvalues $\{\lambda_j\}$ and spectral projectors $\{\Pi_j\}$, this reduces to
\begin{align}
\rho(t)=\sum_j E_\alpha(\lambda_j t^\alpha)\,\Pi_j \rho(0).
\end{align}
For non-diagonalizable $\mathcal{L}$, the same operator expression \eqref{eq:Mittag-Leffler} remains valid and the action on generalized eigenspaces simply involves polynomially weighted derivatives of $E_\alpha$ (Jordan–block structure).  
Notably, the subordination representation and complete monotonicity of $K_\alpha^{(\mathrm{V})}$ do not require diagonalizability.
For $\alpha=1$, one recovers $E_1(z)=e^{z}$ and the familiar exponential Lindblad relaxation. For $0<\alpha<1$, the response crosses over from stretched-exponential at short times to algebraic at long times, capturing persistent memory and non-Markovian relaxation observed in many physical settings~\cite{Metzler2014}.

Because $K_\alpha^{(\mathrm{V})}$ is completely monotone, the dynamical map remains CPTP whenever $\mathcal{L}$ is of GKSL form; however, the resulting family generally lacks the semigroup property and is not CP-divisible~\cite{Rivas2014}. The next subsection makes this precise via the Bochner–Phillips subordination formula, in which physical time is replaced by a random operational time with power-law waiting statistics. Additional details on the Caputo–Volterra equivalence and on fractional Adams–Moulton discretizations appear in Appendices~\ref{app:caputo-volterra} and~\ref{app:adams-moulton}.
Recent analyses of driven two-level systems show analogous fractional relaxation and coherence damping governed by Mittag–Leffler functions, extending the Rabi problem to fractional time evolution~\cite{lopez2025generalisedfractionalrabiproblem}.

\subsubsection{Fractional Subordination}\label{sec:subordination}

Fractional dynamics is equivalent to running the usual Markovian evolution $e^{u\mathcal{L}}$ on a \emph{random clock} $u$ whose statistics encode memory; averaging over that clock yields the fractional map.
\begin{align}
\hat{\rho}(s)=s^{\alpha-1}\big(s^\alpha\mathbb{I}-\mathcal{L}\big)^{-1}\rho(0). \label{eq:frac_map}
\end{align}
Using Bochner’s functional calculus for the GKSL generator $\mathcal{L}$~\cite{Bochner1955,Feller1971,BaeumerMeerschaert2001,Schilling2012Bernstein,MeerschaertSikorskii2012},
\begin{align}
\big(s^\alpha\mathbb{I}-\mathcal{L}\big)^{-1}=\int_0^\infty e^{-u s^\alpha}~e^{u\mathcal{L}}~du, \label{eq:GKSL}
\end{align}
and inverting the Laplace transform introduces the inverse-stable (L\'{e}vy) density $f_\alpha(u,t)$:
\begin{align}
\rho(t)=\int_0^\infty f_\alpha(u,t)~e^{u\mathcal{L}}\rho(0)~du.
\label{eq:subordination}
\end{align}
The kernel satisfies $f_\alpha(u,t)\ge 0$ and $\int_0^\infty f_\alpha(u,t)~du=1$, which makes the evolution~\eqref{eq:subordination} represent a convex mixture of CPTP semigroups \(e^{u\mathcal{L}}\) (detailed in Appendix~\ref{app:subordination}). 
Specifically, the subordination integral in Eq.~\eqref{eq:subordination} represents a \emph{probabilistic construction} of fractional dynamics: the inverse--stable density $f_\alpha(u,t)\!\ge0$ acts as a genuine probability distribution for the random operational time $u$ of an underlying stochastic clock. Consequently, $\Phi_\alpha(t)$ can be interpreted as the statistical mixture $\Phi_\alpha(t)=\mathbb{E}_{U(t)}[e^{U(t)\mathcal{L}}]$, i.e., an ensemble average of standard Markovian (Lindblad) evolutions sampled at random times $U(t)$ with power--law waiting statistics. This probabilistic representation ensures that complete positivity is preserved and explains how algebraic relaxation emerges naturally from the statistics of random operational time. 
 
\textcolor{black}{The fractional order $\alpha$ therefore acquires a clear operational meaning in the subordination picture: it determines the statistics of the random operational time $U(t)$. For an inverse--stable subordinator, the waiting-time distribution obeys $\psi_\alpha(\tau)\sim\tau^{-1-\alpha}$, so that $\alpha$ controls the heaviness of the tail and thus the persistence of memory in the stochastic time change.  When $\alpha=1$, the process reduces to ordinary physical time with exponential waiting statistics, and one recovers the Markovian Lindblad semigroup $\{e^{t\mathcal{L}}\}$.  When $0<\alpha<1$, the broad, power-law waiting-time statistics slow down the operational time and produce non-Markovian relaxation characterized by Mittag–Leffler or power-law decay.  In this sense, $\alpha$ serves as a physically interpretable memory exponent rather than a phenomenological parameter.}

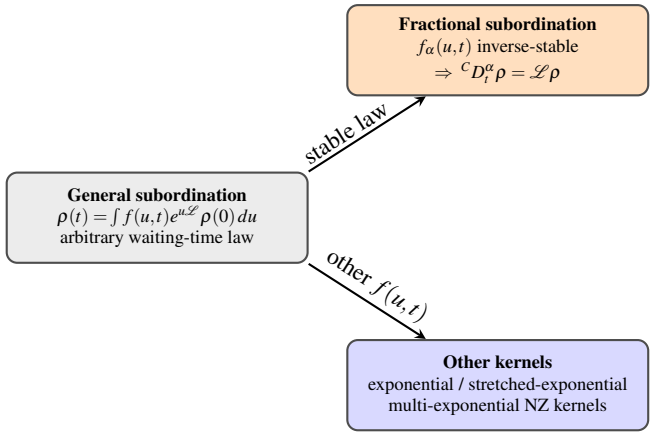
\begin{figure}[t]
\centering
\begin{tikzpicture}[
    node distance=1.0cm,
    box/.style={
        rectangle, rounded corners=4pt,
        draw=black!70, thick,
        align=center,
        minimum width=4cm,
        minimum height=1.2cm,
        font=\scriptsize
    },
    arrow/.style={->, thick, >=stealth}
]

\node[box, fill=gray!15] (general) {
    \textbf{General subordination}\\
    $\rho(t)=\!\int f(u,t)e^{u\mathcal{L}}\rho(0)\,du$\\
    arbitrary waiting-time law
};

\node[box, fill=orange!25, above right=1cm and 0.5cm of general] (frac) {
    \textbf{Fractional subordination}\\
    $f_\alpha(u,t)$ inverse-stable\\[2pt]
    $\Rightarrow~\prescript{C}{}{D}_t^{\alpha}\rho=\mathcal{L}\rho$
};

\node[box, fill=blue!15, below right=1cm and 0.5cm of general] (other) {
    \textbf{Other kernels}\\
    exponential / stretched-exponential\\
    multi-exponential NZ kernels
};

\draw[arrow] (general.north east) -- node[left, xshift = 15pt, yshift=5pt,sloped]{\small stable law} (frac);
\draw[arrow] (general.south east) -- node[above, yshift =-1pt,sloped]{\small other $f(u,t)$} (other);

\end{tikzpicture}
\caption{\textcolor{black}{General subordination replaces physical time by a random operational time. Fractional (inverse–stable) laws yield the fractional subordination, while other waiting-time distributions lead to non-fractional NZ kernels with short- or multi-scale memory (not necessarily power-law).}}
\label{fig:general-subordination}
\end{figure}

\textcolor{black}{
While inverse–stable subordination yields the fractional master equation, it is important to note that the subordination framework itself is far more general (see Figure~\ref{fig:general-subordination}). Any admissible waiting-time density $f(u,t)$ defines a subordinated evolution \eqref{eq:subordination}, and different choices of $f$ generate different classes of non-Markovian memory kernels~\cite{Bochner1955, Feller1971, MeerschaertSikorskii2012}. Fractional (inverse–stable) waiting-time laws are distinguished because they lead to a closed-form Caputo master equation, guarantee complete positivity for any GKSL generator, and produce Mittag–Leffler relaxation with algebraic memory~\cite{Podlubny1999, Mainardi2000, BaeumerMeerschaert2001}. However, subordination with non-stable waiting-time densities, such as exponential, stretched-exponential, or multi-exponential forms, remains mathematically valid and yields NZ-type memory kernels with short-ranged or multi-scale decay rather than power laws~\cite{Breuer2002, Rivas2014, METZLER20001, Metzler_2004}. Thus, fractional subordination should be viewed as a special, analytically tractable instance within a broader class of subordinated Lindblad evolutions: it targets systems whose non-Markovianity is dominated by algebraic or broad temporal correlations, while other waiting-time laws may be used to model qualitatively different memory structures.}

The subordination representation \eqref{eq:subordination} also has an important algorithmic interpretation. Because $e^{u\mathcal{L}}$ admits standard quantum–trajectory and quantum–jump unravelings~\cite{Dalibard1992, Dum1992, Plenio1998}, the subordination formula shows that any non-Markovian trajectory method can be extended to fractional, Markovian dynamics 
\textcolor{black}{by replacing the physical time with a random clock $U(t)$ drawn from $f_\alpha(u,t)$, whose fluctuations become increasingly intermittent as $\alpha$ decreases.  Smaller $\alpha$ thus leads to more strongly non-Markovian trajectory ensembles with longer sojourn periods between effective Lindblad updates.}
Thus, fractional dynamics corresponds to sampling ordinary Lindbladian trajectories at random operational times, with the memory effects arising entirely from the statistics of $U(t)$.

In the special case $\mathcal{L}=-i[H,~\cdot~]$, Eq.~\eqref{eq:subordination} reduces to a fractional Liouville evolution,
\begin{align}
\rho(t)
   = \!\int_0^{\infty}\! f_{\alpha}(u,t)\,
      e^{-iHu}\rho(0)e^{iHu}\,du ,
\end{align}
which represents a convex mixture of unitary maps parameterized by the random operational time~$u$. Because $f_{\alpha}(u,t)$ is non-negative and normalized, this evolution is CPTP for all~$\alpha$, continuously connecting the coherent, memoryful fractional regime to the strictly unitary Liouville dynamics recovered in the limit $\alpha\!\to\!1$. This establishes the Liouville equation as the coherent base of the hierarchy spanning unitary, Markovian, and non-Markovian quantum evolution.

Note that, one can plug Eq.~\eqref{eq:GKSL} into Eq.~\eqref{eq:frac_map}, where the obtained Laplace kernel \(s^{\alpha-1}e^{-us^{\alpha}}\) is completely monotone, implying that \(f_\alpha(u,t)\) is a probability density on \(u\ge0\).
Consequently, the subordinated map 
\begin{align}
\Phi_\alpha(t):\ \rho(0)\mapsto \rho(t)
\end{align}
preserves complete positivity and trace, i.e., remains CPTP whenever the underlying generator \(\mathcal{L}\) is of GKSL form~\cite{Bochner1955,Feller1971,Hall2014,Chruscinski2017}. In standard (Markovian) open dynamics, the family $\{\Phi_1(t)\}_{t\ge 0}=\{e^{t\mathcal{L}}\}$ forms a semigroup obeying the composition law, Eq.~\eqref{eq:composition}; equivalently, for any $0\le \tau\le t$ there exists a CPTP map $\Lambda(t,\tau)$ such that
\begin{align}
\Phi_1(t)=\Lambda(t,\tau)~\Phi_1(\tau)\qquad\text{(CP-divisibility)}.
\end{align}
By contrast, the fractional dynamics ($0<\alpha<1$) averages over random clock times in Eq.~\eqref{eq:subordination} and thereby breaks exact time composition in general:
\begin{align}
\Phi_\alpha(t)\neq \Phi_\alpha(t-\tau)~\Phi_\alpha(\tau),
\qquad 0\le \tau < t,
\end{align}
so $\{\Phi_\alpha(t)\}$ is typically not a semigroup. A simple test follows from the known action on eigenoperators of $\mathcal{L}$:
\begin{align}
\mathcal{L}X=\lambda X \ \Rightarrow\ \Phi_\alpha(t)X = E_\alpha(\lambda t^\alpha)~X.
\end{align}
If $\{\Phi_\alpha(t)\}$ were a semigroup, one would require $E_\alpha(\lambda t^\alpha)=e^{\mu t}$ for all $t$, which is impossible unless $\alpha=1$ (or $\lambda=0$). Hence, for $0<\alpha<1$ and nontrivial $\mathcal{L}$, the fractional family is not a semigroup and is generally not CP-divisible. In special commuting cases (e.g. pure dephasing channels) where all maps diagonalize simultaneously, intermediate CPTP maps may still exist, see, for example, Refs.~\citenum{PhysRevLett.112.120404,Sandev_2019,Ang2024MemoryInducedWeakDissipation}).

For $\lambda<0$, the Mittag--Leffler factor obeys $E_\alpha(\lambda t^\alpha)\sim -\big[\lambda~\Gamma(1-\alpha)\big]^{-1} t^{-\alpha}$ as $t\to\infty$. Populations and coherences therefore relax as a power law $t^{-\alpha}$ rather than exponentially, an experimentally relevant signature of long memory~\cite{Weiss2012,Mainardi2000}. In short, subordination guarantees CPTP at each time $t$ while replacing the Markovian semigroup structure with a memory-induced, non-composable flow that exhibits algebraic relaxation.

\begin{figure*}
\centering
\begin{tikzpicture}[
    node distance=1.2cm,
    box/.style={
        rectangle, rounded corners=6pt,
        draw=black!80, thick,
        align=center,
        minimum width=4.8cm,
        minimum height=1.5cm,
        font=\small
    },
    arrow/.style={->, thick, >=stealth}
]

\node[box, fill=orange!25] (frac) {
    \textbf{Fractional Lindblad} \\
    $\prescript{C}{}{D}_t^\alpha \rho = \mathcal{L}\rho$ \\
    Mittag--Leffler modes, long-memory
};

\node[box, fill=purple!15, above left=1.2cm and 1.2cm of frac] (nz) {
    \textbf{NZ QME} \\
    Laplace kernel: $\tilde{\kappa}_{\mathrm{NZ},\alpha}(s)=s^{1-\alpha}\mathcal{L}$ \\
    $(s - \tilde{\kappa}_{\mathrm{NZ},\alpha}(s))^{-1}$
};

\node[box, fill=blue!15, above right=1.2cm and 1.2cm of frac] (heom) {
    \textbf{HEOM} \\
    Self-energy: $\Sigma(s)\sim s^{\chi}$ \\
    $(s^\alpha\mathbb{I}-\mathcal{L})^{-1}$,~~$\alpha=\chi+1$
};

\node[box, fill=green!20, below left=1.2cm and 1.2cm of frac] (sub) {
    \textbf{Subordination / CTRW} \\
    $\rho(t)=\displaystyle\int_0^\infty f_\alpha(t,u)\,e^{u\mathcal{L}}\rho(0)\,du$ \\
    L\'{e}vy-time randomization
};

\node[box, fill=yellow!25, below right=1.2cm and 1.2cm of frac] (if) {
    \textbf{Influence functional}\\
    QUAPI, MCTDH, path integrals\\
    explicit bath correlations
};

\draw[arrow] (nz.south east) -- node[above, xshift = -7pt, yshift=-2pt, sloped]{\small memory-kernel} (frac);
\draw[arrow] (nz.south east) -- node[above, xshift = -8pt, yshift=-12pt, sloped]{\small deformation}(frac);
\draw[arrow] (heom.south west) -- node[above, xshift=10pt,yshift=-2pt, sloped]{\small self-energy} (frac);
\draw[arrow] (heom.south west) -- node[above, xshift=10pt,yshift=-12pt, sloped]{\small deformation} (frac);
\draw[arrow] (sub.north east) -- node[right, xshift=-40pt, yshift=5pt, sloped]{\small operational time} (frac);
\draw[arrow] (sub.north east) -- node[right, xshift=-35pt, yshift=-5pt, sloped]{\small randomization} (frac);
\draw[arrow] (if.north west) -- node[above, xshift=5pt, yshift=-2pt,sloped]{\small coarse-graining} (frac);
\draw[arrow] (if.north west) -- node[above, xshift=5pt, yshift=-13pt,sloped]{\small of long memory} (frac);

\end{tikzpicture}

\caption{\textcolor{black}{Algebraic connections between the fractional Lindblad equation and mainstream non--Markovian approaches. Fractional dynamics arise as (i) a resolvent-level deformation of the NZ memory-kernel equation, (ii) a coarse-grained surrogate for HEOM self-energy structures, (iii) a subordinated Markovian process corresponding to L\'{e}vy-distributed operational times, and (iv) a compact representation of long-memory features also encoded in influence-functional methods such as QUAPI and MCTDH.}}
\label{fig:connection-nonmarkov}
\end{figure*}
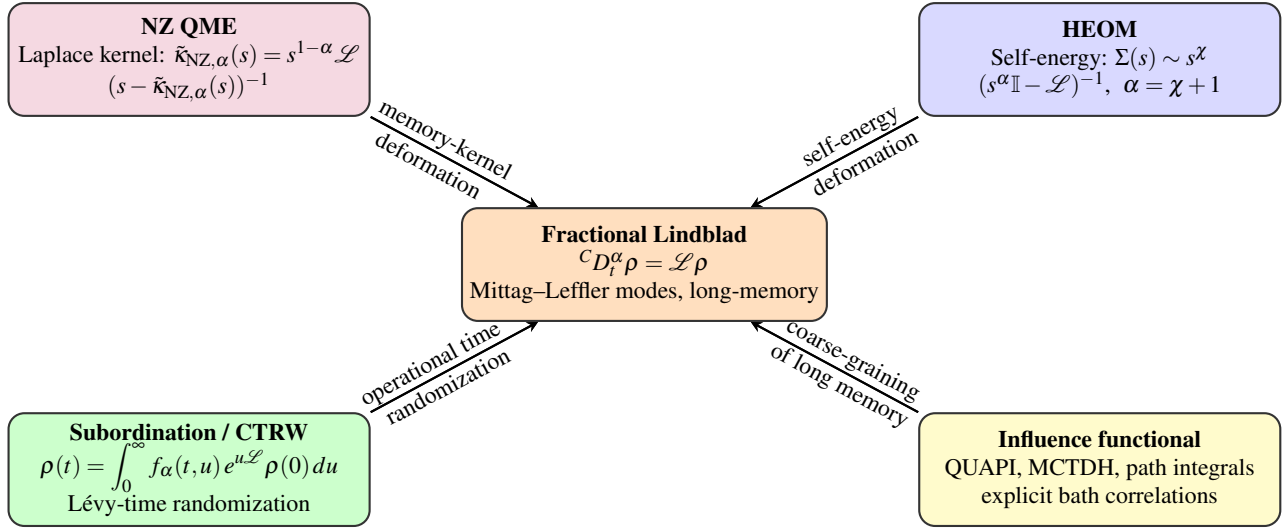

\subsubsection{Algebraic connections to Nakajima--Zwanzig memory kernels and hierarchical equations of motion}
\label{subsec:connections}

\textcolor{black}{
The fractional master equation admits a unified algebraic structure that connects it directly to several mainstream non-Markovian formalisms. In particular, the Laplace-transform solution of the Caputo equation, Eq.~\eqref{eq:frac_map}, shows that fractional dynamics correspond to a \emph{spectral deformation} of the standard Lindbladian resolvent $(s\mathbb{I}-\mathcal{L})^{-1}$ via the substitution $s \mapsto s^\alpha$. As discussed in the previous subsection, this deformation underlies the quantum analogue of subordinated Markov processes and continuous-time random walks (CTRWs)~\cite{Bochner1955, Feller1971, BaeumerMeerschaert2001, Schilling2012Bernstein, MeerschaertSikorskii2012, Metzler_2004, METZLER20001}, where heavy-tailed waiting-time statistics generate non-Markovian relaxation. We now elucidate how the same spectral deformation relates algebraically to other major approaches, including NZ projection-operator kernels, HEOM, and influence-functional methods.}

\vspace{0.15cm}

\textcolor{black}{
\paragraph{Connection to NZ memory kernels.}
A general NZ equation takes the convolution form~\cite{Nakajima1958,Zwanzig1960}
\begin{align}
\dot{\rho}(t) = \int_0^t \kappa_{\rm NZ}(t-\tau)\,\rho(\tau)\,d\tau,
\end{align}
with Laplace-domain solution
\begin{align}
\tilde{\rho}(s) = (s\mathbb{I} - \tilde{\kappa}_{\rm NZ}(s))^{-1}\rho(0).
\end{align}
Comparing this with the fractional resolvent~\eqref{eq:frac_map} identifies the NZ kernel that reproduces fractional dynamics:
\begin{align}
\tilde{\kappa}_{{\rm NZ},\alpha}(s) = s^{1-\alpha}\mathcal{L}.
\end{align}
In the time domain this corresponds to a weakly singular power-law kernel $\kappa_{{\rm NZ},\alpha}(t)=t^{\alpha-2}\mathcal{L}/\Gamma(\alpha-1)$, demonstrating that the fractional master equation is algebraically equivalent to a non-Markovian NZ equation with a completely monotone memory kernel.}

\vspace{0.15cm}

\textcolor{black}{
\paragraph{Connection to HEOM.}
In HEOM~\cite{HEOM_1,HEOM_2,HEOM_3,HEOM_4}, Gaussian environments generate a frequency-dependent self-energy $\Sigma(s)$ such that
\begin{align}
\tilde{\rho}(s) = (s\mathbb{I} - \mathcal{L} - \Sigma(s))^{-1}\rho(0).
\end{align}
For structured spectral densities with low-frequency scaling $J(\omega)\sim \omega^\chi$, the bath correlation function acquires a power-law tail and the associated self-energy displays the asymptotic scaling $\Sigma(s)\sim s^\chi$. In this regime, the HEOM resolvent becomes algebraically equivalent to $(s^\alpha\mathbb{I}-\mathcal{L})^{-1}$ with $\alpha=\chi+1$, up to prefactors.}

\vspace{0.15cm}
\textcolor{black}{
\paragraph{Connection to influence-functional approaches.}
Path-integral methods such as QUAPI~\cite{Makri1995QUAPI1, Makri1995QUAPI2}, MCTDH-based open-system propagation~\cite{Thoss2001MCTDH, Wang2003MCTDHReview}, and related influence-functional techniques provide numerically controlled descriptions of non-Markovian dynamics by explicitly encoding bath correlation functions in discretized memory tensors or augmented Hilbert spaces. These approaches do not yield closed-form propagators and often scale exponentially with the bath memory time. The fractional master equation does not aim to reproduce their microscopic structure; instead, it serves as a compact surrogate for environments whose memory exhibits broad temporal support or algebraic tails, thereby complementing these methods at a coarse-grained level.
}

\vspace{0.15cm}

\textcolor{black}{
Together, these algebraic correspondences clarify the generality of fractional dynamics. For any GKSL generator, each Lindbladian eigenmode acquires a Mittag--Leffler relaxation profile, and the deformation $s\mapsto s^\alpha$ modifies the spectral response of the resolvent without altering the dissipator. Fractional dynamics therefore provide a low-parameter, CPTP surrogate that captures long-memory behavior reflected in NZ kernels, HEOM self-energies, and influence-functional approaches, while remaining algebraically consistent with established open-system theories.
}


\section{Numerical Demonstration and Discussion}
\label{sec:numerical-discussion}

\textcolor{black}{To assess the physical relevance of the fractional Liouville–Lindblad dynamics, we benchmark it against a well-understood non-Markovian system for which an exact microscopic solution is available. The pure-dephasing spin–boson model provides an ideal testbed: its coherence dynamics are governed entirely by bath-induced memory effects, and the off-diagonal elements of the reduced density matrix admit a closed-form expression determined by the spectral density. This enables a precise, parameter-free comparison between (i) the exact coherence function derived from the full system–bath Hamiltonian (i.e. the ground-truth) and (ii) the coherence predicted by the fractional evolution obtained through subordination of a Lindblad generator, without numerical or model-dependent ambiguities. Our goal in this section is to evaluate whether the fractional propagation captures the characteristic non-exponential decay, long-memory behavior, and turnover between Gaussian and exponential regimes exhibited by the microscopic model. Establishing this agreement demonstrates that the fractional dynamics can faithfully reproduce structured non-Markovian behavior without requiring explicit microscopic modeling of the environment.}

\begin{figure*}[ht]
    \centering
    \includegraphics[width=0.95\linewidth]{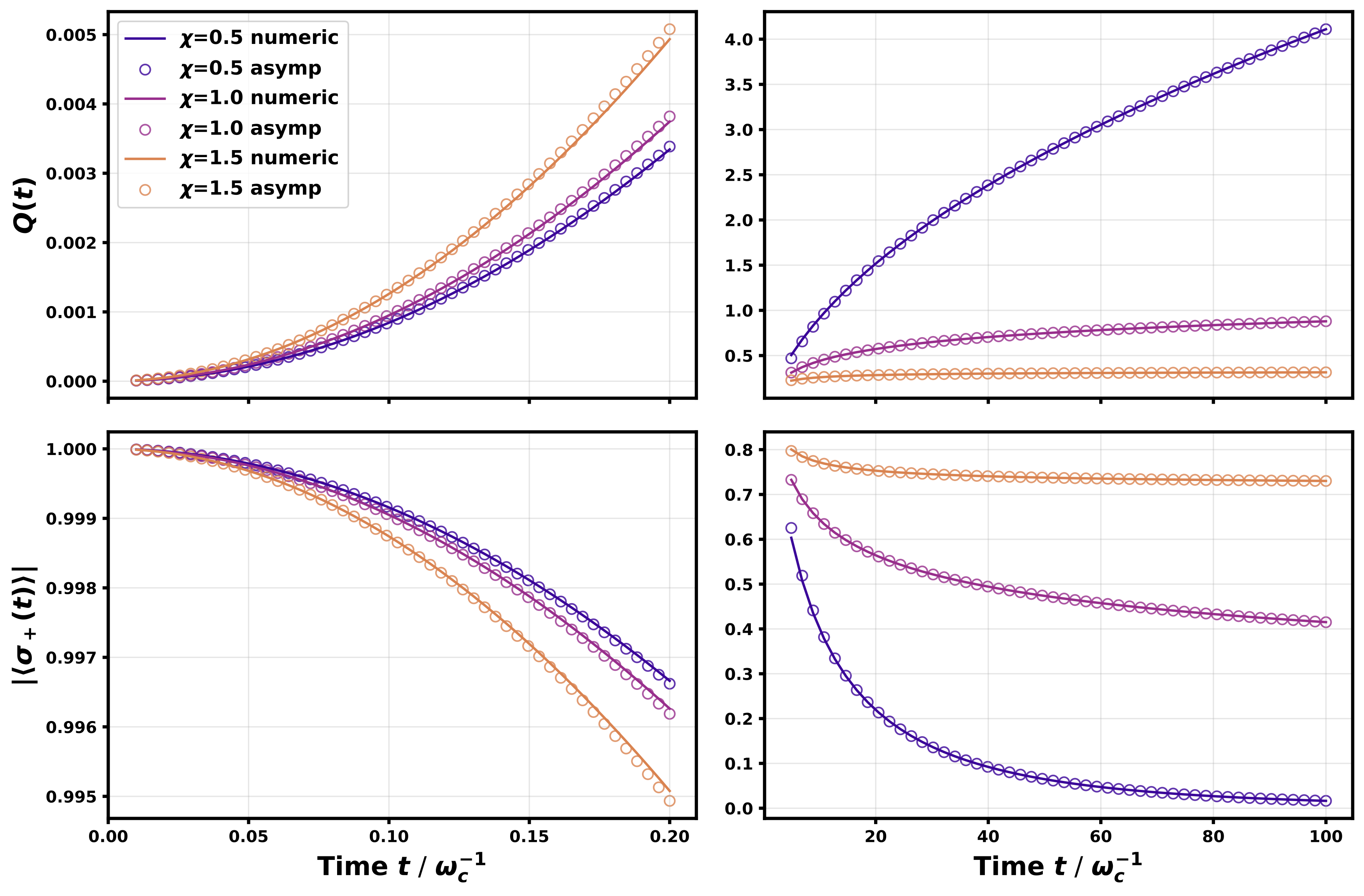}
    \caption{Short-time ($t \le 0.2/ \omega_c$, left panels) and long-time ($t \ge 5/\omega_c$, right panels) asymptotic propagation of the dephasing functional $Q(t)$ and corresponding coherence amplitude $|\langle\sigma_+(t)\rangle|$ for three different bath exponents $\chi=0.5$, $1.0$, and $1.5$. Solid lines denote numerical integrations of Eq.~\eqref{eq:Q-general}; open circles show asymptotic expressions (Table.~\ref{tab:Q-short-long}) with optimized prefactors, consistent with the asymptotic analysis in Refs.~\citenum{Leggett1987,Grifoni1998,Weiss2012,deVega2017}.}
    \label{fig:Q_Sigma_exact}
\end{figure*}

\textcolor{black}{For this purpose, we now summarize the exact coherence dynamics of the pure-dephasing spin–boson model, which will serve as the microscopic reference against which the fractional evolution is compared.} We work in the exactly solvable pure--dephasing limit of the spin--boson model,
\begin{align}
H = \frac{\epsilon}{2}~\sigma_z
 + \sum_k \omega_k~ b_k^\dagger b_k
 + \sigma_z \sum_k g_k ~(b_k^\dagger + b_k),
\label{eq:sbm-ham}
\end{align}
with independent boson modes $\{b_k\}$ in thermal equilibrium at inverse temperature $\beta$. Since $[H,\sigma_z]=0$, populations are constants of motion; only the off--diagonal element (coherence) decays.
If we assume a factorized state for the total spin-boson system
\begin{align}
    \rho_{\text{tot}}(0) = \rho_S(0) \otimes \rho_B,
\end{align}
with bath state $\rho_B = \frac{e^{-\beta H_B}}{Z_B}$ and bath Hamiltonian $H_B = \sum_k \omega_k b_k^\dagger b_k$. 
Since $\rho_B$ is exactly a Gaussian state, tracing over a thermal Gaussian bath yields the exact coherence (Kubo--Martin--Schwinger identity)~\cite{Leggett1987,Weiss2012,deVega2017},
\begin{align}
u(t) = \langle \sigma_+(t)\rangle 
= \langle \sigma_+(0)\rangle ~ e^{-i\epsilon t}~ e^{-Q(t)}, \label{eq:Gaussian_bath}
\end{align}
where $\sigma_+ = \tfrac{1}{2}(\sigma_x+i\sigma_y)$ and 
\begin{align}
Q(t) = \frac{2}{\pi}\int_0^\infty d\omega~ \frac{J(\omega)}{\omega^2}~
\Big[ (1-\cos\omega t) \coth \Big(\frac{\beta \omega}{2}\Big)\Big]
\label{eq:Q-general}
\end{align}
with $J(\omega)$ the spectral density. A standard choice of $J(\omega)$ is the algebraic (sub--Ohmic/Ohmic/super--Ohmic) family with exponential cutoff~\cite{Breuer2002,Weiss2012},
\begin{align}
J(\omega) = \eta~ \omega^\chi \omega_c^{~1-\chi} e^{-\omega/\omega_c},
\quad \chi\in\mathbb{R},~\eta>0,~\omega_c>0.
\label{eq:J-omega}
\end{align}
as used in standard treatments of dissipative two-level systems and spin–boson dynamics~\cite{Grifoni1998,Weiss2012}.
Table~\ref{tab:Q-short-long} provides the asymptotic scaling for the coherence that smoothly interpolates between Gaussian short-time decay, stretched-exponential-like intermediate behavior, and algebraic or saturating long-time tails, depending on $\chi$ and $t$~\cite{Leggett1987,Grifoni1998,Weiss2012,deVega2017}. The real-time propagation of the dephasing functional $Q(t)$ and corresponding coherence amplitude $|\langle\sigma_+(t)\rangle|$ for $t\le 0.2/\omega_c$ and $t\ge 5/\omega_c$ and three different bath exponents $\chi=0.5$, $1.0$, and $1.5$ are shown in Figure~\ref{fig:Q_Sigma_exact}. 
In the weak-coupling and pure-dephasing limit, the reduced coherence $u(t)$ obeys a NZ convolution equation,
\begin{align}
\dot{u}_{\mathrm{NZ}}(t)=-\int_0^t\kappa_{\mathrm{NZ}}(t-\tau) u_{\mathrm{NZ}}(\tau) d\tau,
\label{eq:nz-scalar}
\end{align}
where the memory kernel $\kappa_{\mathrm{NZ}}(t)$ is determined by the bath correlation function
\begin{align}
\kappa_{\rm NZ} \propto C(t)&=\langle B(t)B(0)\rangle \notag \\
&= \frac{2}{\pi}\int_0^\infty d\omega~ J(\omega)
 \cos(\omega t) \coth \Big(\frac{\beta \omega}{2}\Big) \label{eq:NZ_kernel}
\end{align}
with 
\begin{align}
    B(t)=\sum_k g_k\big(b_k^\dagger e^{i\omega_k t}+b_k e^{-i\omega_k t}\big),
\end{align}
corresponding to the bath coupling operator.
The exact solution,
\begin{align}
u_{\mathrm{NZ}}(t)=e^{-Q(t)}. \label{eq:NZ_expQ}
\end{align}
coincides with the Gaussian-bath result in Eq.~\eqref{eq:Gaussian_bath}, confirming that $Q(t)$ fully characterizes the non-Markovian dephasing kernel. \textcolor{black}{In the following sections, Eq.~\eqref{eq:NZ_expQ} serves as the baseline against which the conventional and fractional evolutions are evaluated.}

\begin{table*}[ht]
\centering
\caption{Short- and long-time asymptotic forms of the dephasing functional $Q(t)$ and corresponding coherence amplitude $|\langle\sigma_+(t)\rangle|$ for different bath exponents $s$. Here, the asymptotic forms in each column follow standard analyses of the spin–boson model, and focus on the non-thermal quantum decoherence, thus $\coth(\beta \omega/2)\sim 1$. For $|\langle\sigma_+(t)\rangle|$ in short-time Gaussian decay, $t_G=\sqrt{2}~\omega_c^{-1} (\eta\,\Gamma(\chi+1))^{-1/2}$. For $|\langle\sigma_+(t)\rangle|$ in long-time sub-ohmic case, $C_\chi(\omega_c)=\frac{2}{\pi}\,\eta\,\omega_c^{\,1-\chi}\Gamma(1-\chi)\sin\!\Big(\frac{\pi \chi}{2}\Big)$. For $|\langle\sigma_+(t)\rangle|$ in the long-time super-Ohmic case, $Q_\infty=\tfrac{2}{\pi}\eta\,\Gamma(\chi\!-\!1)$.}
\label{tab:Q-short-long}
\renewcommand{\arraystretch}{1.3}
\setlength{\tabcolsep}{6pt}
\begin{ruledtabular}
\begin{tabular}{lllll}
 & Short-time~\cite{Leggett1987,Loss1998,Weiss2012} & \multicolumn{3}{c}{Long-time ($t \gg 1/\omega_c$)} \\
\cline{3-5}
 & ($t \ll 1/\omega_c$) & Sub--Ohmic ($\chi<1$)~\cite{Leggett1987,Grifoni1998} & Ohmic ($\chi=1$)~\cite{Leggett1987,Weiss2012,deVega2017} & Super--Ohmic ($\chi>1$)~\cite{Grifoni1998,Weiss2012} \\
\hline
$Q(t)$ &
$\displaystyle \mathcal{O}\left(\frac{1}{2}\eta\,\Gamma(\chi+1)\,\omega_c^2\,t^2\right)$ &
$\displaystyle \mathcal{O}\left(C_\chi(\omega_c)\,t^{1-\chi}\right)$ &
$\displaystyle \mathcal{O}\left(\eta\ln(\omega_c t)\right)$ &
$\displaystyle \mathcal{O}\left(Q_\infty - D_\chi\,t^{-(\chi-1)}\right)$ \\[6pt]
$|\langle\sigma_+(t)\rangle|$ &
$\displaystyle \mathcal{O}\left(e^{-t^2/t_G^2}\right)$ &
$\displaystyle \mathcal{O}\left(e^{-\omega_c^{1-\chi}\,t^{1-\chi}}\right)$ &
$\displaystyle \mathcal{O}\left(t^{-\eta}\right)$ &
$\displaystyle \mathcal{O}\left(e^{-Q_\infty}\!\big[1 + D_\chi\,t^{-(\chi-1)}\big]\right)$ 
\end{tabular}
\end{ruledtabular}
\end{table*}

\subsection{Why Markovian dephasing misses $e^{-Q(t)}$?}
\label{sec:lindblad-vs-nz}

\begin{figure*}[ht]
    \centering
    \includegraphics[width=0.95\linewidth]{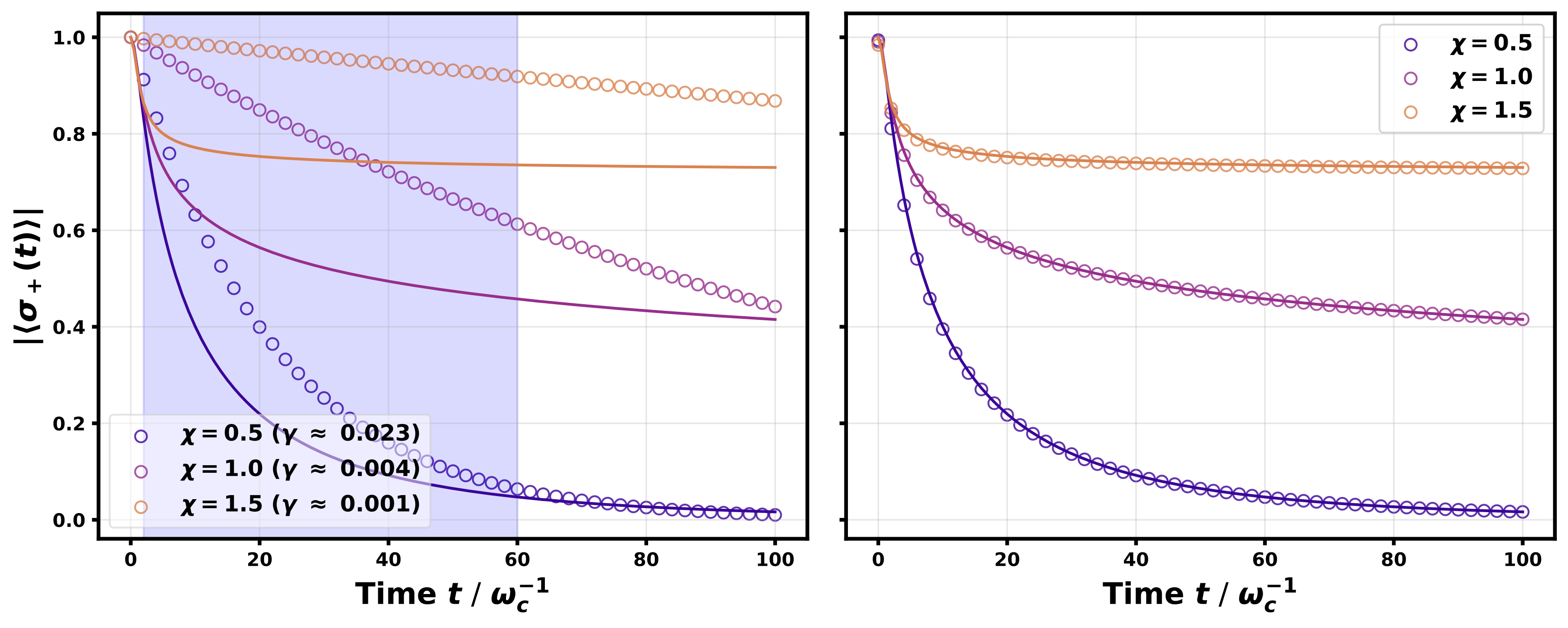}
    \caption{Comparison of exact coherence from Figure~\ref{fig:Q_Sigma_exact} with Markovian and non-Markovian fits. The asymptotic propagation of the coherence amplitude $|\langle\sigma_+(t)\rangle|$ for three different bath exponents $\chi=0.5$, $1.0$, and $1.5$ are computed by constant $\gamma$ (Markovian, left panel) and time-local $\gamma(t) = \tfrac{1}{2}\dot{Q}(t)$ (non-Markovian, right panel), respectively. The constant rate $\gamma$ is obtained by fitting $\log|\langle\sigma_+(t)\rangle|$ to a linear decay in the intermediate-time regime $t\in [2/\omega_c, 60/\omega_c]$ (blue shade, left panel), while solid lines are obtained by numerically integrating Eq.~\eqref{eq:Q-general}. \textcolor{black}{The parameters shown here ($\gamma$ and $\gamma(t)$) belong to Markovian and time-local models only and are not related to the fractional parameters $(\alpha,\lambda)$ introduced in Figure~\ref{fig:frac_general}.
}}
    \label{fig:Markovian}
\end{figure*}

A common Markovian dephasing of the spin-boson model relies on the time‑homogeneous Lindbladian model~\eqref{eq:Lindblad1} and \eqref{eq:Lindblad2} with $\gamma_j \ge 0$. Here, the system Hamiltonian $H_S = \frac{\epsilon}{2}\sigma_z$ and a single jump operator $L=\sigma_z$ is employed. This Markovian model can then be transformed to
\begin{align}
\dot{u}_{\rm M}(t) = (i\epsilon - 2\gamma)\,u_{\rm M}(t) \label{eq:markovian_model}
\end{align}
that yields an exponential dephasing
\begin{align}
u_{\rm M}(t) = u_{\rm M}(0)\,e^{i\epsilon t}\,e^{-2\gamma t}.
\label{eq:markov-dephasing}
\end{align}
Eq.~\eqref{eq:markovian_model} can also be written in the NZ form 
\begin{align}
\dot{u}_{\rm M}(t)=-\int_0^t \kappa_{\rm M}(t-\tau)\,u_{\rm M}(\tau)\,d\tau
\label{eq:nz-delta}
\end{align}
with a $\delta$-kernel,
\begin{align}
    \kappa_{\rm M}(t)=2\gamma \delta(t).
\end{align}
Since the entire Markovian kernel is a $\delta$-function concentrated at the present time, this is different from the exact kernel~\eqref{eq:NZ_kernel} that possesses a long tail inherited from the bath spectrum $J(\omega)$. Therefore, the time-homogeneous Lindbladian cannot capture the long-time memory that produces $u(t)=e^{-Q(t)}$. 
\textcolor{black}{The left panel of Figure~\ref{fig:Markovian} explicitly illustrates the discrepancy, where a constant Markovian rate~$\gamma$ produces a purely exponential decay that systematically deviates from the exact coherence $u(t)=e^{-Q(t)}$}. 
The ``missing terms'' are precisely the nonlocal (finite-memory) contributions.
It is worth noting that, in order to reproduce $u(t)=e^{-Q(t)}$, one can choose
the rate $\gamma$ to be time-dependent. For example, with $\gamma(t)=\tfrac12\dot{Q}(t)$, we have
\begin{widetext}
\begin{align}
\dot{u}(t) = \big(i\epsilon - 2\gamma(t)\big)\,u(t)
~~\Rightarrow~~
    u(t)=u(0)\,e^{i\epsilon t}\,\exp\!\Big(-2\!\int_0^t\!\gamma(\tau)\,d \tau\Big).
\label{eq:tcl}
\end{align}
\end{widetext}
\textcolor{black}{Numerically, as can be seen from the right panel of Figure~\ref{fig:Markovian}, a time-local rate $\gamma(t)=\tfrac{1}{2}\dot{Q}(t)$ can reproduce the exact behavior only by encoding the full non-Markovian memory directly into the time-dependent rate.}
This more flexible time-local dephasing model merely re‑encodes the exact memory in $\gamma(t)$ through $Q(t)$ to reproduce the exact dephasing law. 
Moreover, if $\gamma(t)$ were to become temporarily negative (possible for structured environments), the dynamics is non-Markovian (not CP-divisible)~\cite{Rivas2014,Hall2014}, and the GKSL form with constant positive rates is no longer valid.

\textcolor{black}{We emphasize that the quantities fitted in Figure~\ref{fig:Markovian} (the constant rate~$\gamma$ or the time-local rate~$\gamma(t)$) pertain solely to the Markovian and time-local Lindblad descriptions. The associated analysis highlights that a constant-rate Lindbladian cannot capture the long-tailed memory encoded in \(Q(t)\), while a time-dependent rate merely re-parametrizes this memory without providing structural insight. The purpose of Figure~\ref{fig:Markovian} is therefore to demonstrate the limitations of these Lindbladian descriptions and to motivate the introduction of a more fundamental generalization that embeds the memory directly into the generator. We next show that replacing the first-order derivative in the Lindblad equation with a fractional derivative produces a compact, CPTP, and physically transparent model that reproduces the exact non-Markovian coherence \(e^{-Q(t)}\) across spectral regimes. Remarkably, the parameters used in the Lindbladian descriptions in this section are not related to the fractional model introduced in the next section and do not involve the fractional order~$\alpha$. }


\begin{figure*}[ht]
    \centering
    \includegraphics[width=\linewidth]{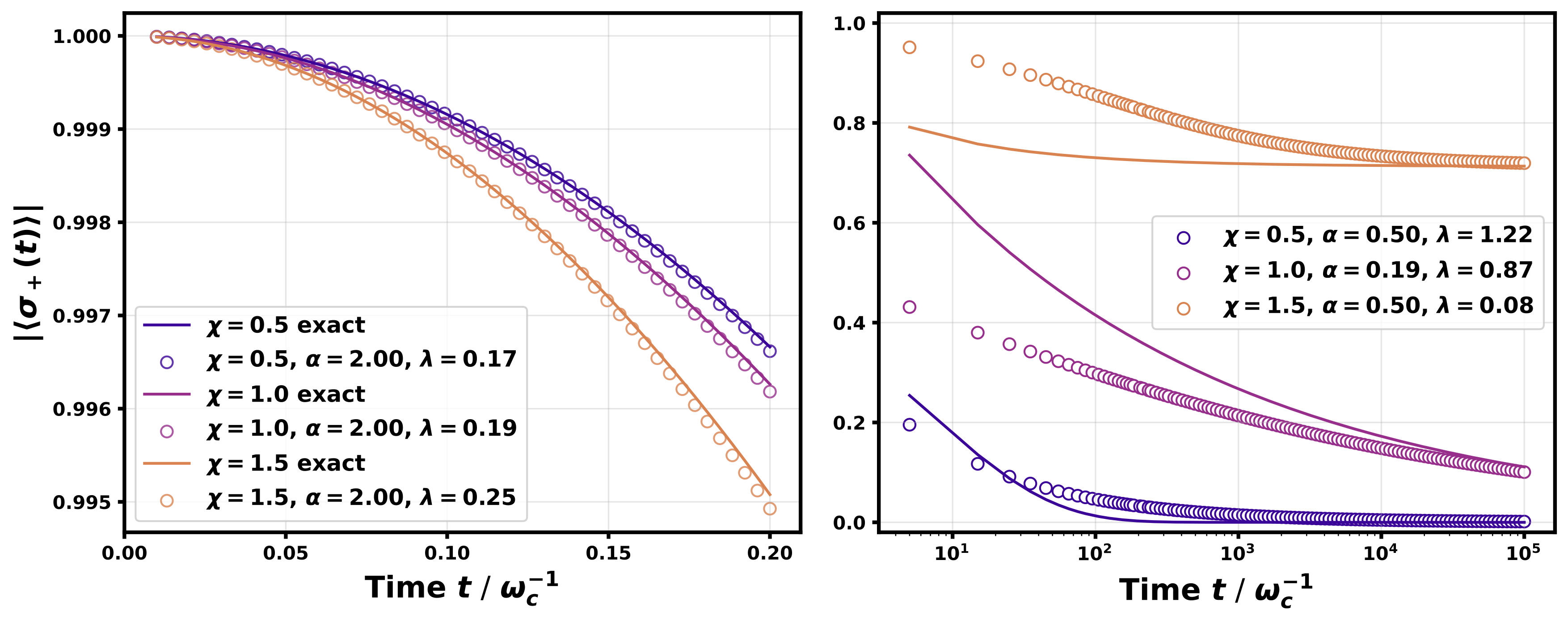}
    \caption{\textcolor{black}{Comparison between the exact coherence $u(t)=e^{-Q(t)}$ (solid lines) and the parameter–free fractional predictions $u_\alpha(t)=E_\alpha(-\lambda t^\alpha)$ (circles) for sub–Ohmic ($\chi=0.5$), Ohmic ($\chi=1$), and super–Ohmic ($\chi=1.5$) baths. The fractional parameters $(\alpha_{\rm pred},\lambda_{\rm pred})$ used here are obtained \emph{directly} from microscopic information: (Left) $\alpha=2$ and $\lambda=2A_\chi$ follow from the universal short-time Gaussian behavior of $Q(t)$ ($t<0.2/\omega_c$); (Right) the long-time values of $(\alpha_{\rm pred},\lambda_{\rm pred})$ follow from the asymptotic structure of $Q(t)$ determined by the spectral density $J(\omega)\propto \omega^\chi$ ($5/\omega_c \le t \le 10^{5}/\omega_c$). The analytic fractional model reproduces both the correct short-time curvature and the correct long-time decay or plateau behavior for all bath types without any fitting. As expected, the discrepancies appear in the intermediate-time crossover regime, where neither the Gaussian expansion nor the asymptotic form of $Q(t)$ is accurate; this region reflects subleading features of the true spectral density and may be used to refine or infer additional structure in $J(\omega)$.}}
    \label{fig:no_fitting_sub}
\end{figure*}

\subsection{Single--order fractional Lindbladian for structured baths}
\label{sec:frac-lindbladian}

The failure of the conventional Markovian Lindbladian to reproduce the nonexponential coherence $e^{-Q(t)}$ underscores the need for a minimal generalization that can capture the memory effects introduced by a structured environment. A natural extension is to replace the first--order time derivative in the Lindblad equation with a \emph{fractional} derivative, producing a non--Markovian but still completely positive and trace--preserving (CPTP) evolution.

We consider the stationary fractional equation of motion
\begin{align}
\label{eq:frac-lindblad}
\prescript{C}{}{D}_t^{\alpha}\rho(t)
   = \lambda\,\mathcal{L}\rho(t), \qquad 0<\alpha<1 ,
\end{align}
where $\mathcal{L}$ is the standard Lindbladian generator. Note that Eq.~\eqref{eq:frac-lindblad} provides the formal CPTP generator: the fractional evolution is a \emph{subordination} of the usual semigroup $e^{u\mathcal{L}}$ over an inverse--stable operational--time distribution~\cite{Mainardi2000,Tarasov2021}.
For pure dephasing, $\mathcal{L}$ acts diagonally in the energy basis, and the off--diagonal element $u(t)=\rho_{01}(t)$ satisfies
\begin{align}
\label{eq:scalar-frac}
\prescript{C}{}{D}_t^{\alpha}u(t)=-\lambda u(t),\qquad u(0)=1 .
\end{align}
Taking Laplace transforms on both sides gives
\begin{align}
s^{\alpha}\tilde u(s)-s^{\alpha-1}=-\lambda\tilde u(s)
~~\Rightarrow~~ \tilde u(s)=s^{\alpha-1}/(s^{\alpha}+\lambda),
\end{align}
whose inverse transform yields the Mittag--Leffler relaxation,
\begin{align}
u_\alpha(t)=E_{\alpha}(-\lambda t^{\alpha}) ,
\label{eq:ML}
\end{align}
where $(\alpha,\lambda)$ do not correspond to the Markovian or time-local rates $(\gamma,\gamma(t))$ discussed previously, and can be found for reproducing the expected dephasing.
The same Mittag–Leffler envelope has recently appeared in the context of the generalized fractional Rabi problem, where fractional differentiation modulates the coherent oscillations of a driven two-level system~\cite{lopez2025generalisedfractionalrabiproblem}.

\textcolor{black}{The rest of this section is organized as follows:  
(i)~a physics-based recipe for predicting $(\alpha,\lambda)$ from the spectral density;
(ii)~an optional refinement scheme (least-squares fitting) used only for benchmarking 
against the exactly solvable spin–boson model; and  
(iii)~a dedicated super–Ohmic ansatz required when the coherence saturates at a finite plateau.}

\begin{table}
\centering
\caption{\textcolor{black}{Predictive fractional parameters $(\alpha_{\rm pred},\lambda_{\rm pred})$ obtained directly from microscopic inputs. The universal short-time parameters follow from the Gaussian expansion of $Q(t)$, while the long-time parameters follow from the asymptotic form of $Q(t)$ determined by the bath exponent $\chi$. All derivations are given in Appendix~\ref{app:alpha-lambda-derivation}.}}
\label{tab:alpha-lambda-recipe}
\resizebox{\linewidth}{!}{
\begin{tabular}{lclclcl}
\hline\hline
&&&&&&\\[-0.2cm]
Regime / Bath type && Structure of $Q(t)$ &&
$\alpha_{\rm pred}$ && \qquad$\lambda_{\rm pred}$ \\[-0.2cm]
&&&&&&\\
\hline
&&&&&&\\[-0.25cm]

\textbf{Universal short time}
&& $A_\chi\, t^2$
&& $2$
&& \qquad $2A_\chi$ \footnote{$A_\chi=\eta\,\Gamma(\chi+1)/\pi$}
\\[6pt]

\hline
&&&&&&\\[-0.25cm]

\textbf{Sub–Ohmic} ($0<\chi<1$)
&& $C_\chi\, t^{1-\chi}$
&& $1-\chi$
&& \qquad $C_\chi \Gamma(2-\chi)$
\\[4pt]

\textbf{Ohmic} ($\chi=1$)
&& $\tfrac{2\eta}{\pi}\,\ln t$
&& $\eta$
&& \qquad $1/\Gamma(1-\eta)$
\\[4pt]

\textbf{Super–Ohmic} ($\chi>1$)
&& $Q_\infty - D_\chi\, t^{-(\chi-1)}$
&& $\chi-1$
&& \qquad $\frac{e^{Q_\infty}-1}{D_\chi\Gamma(2-\chi)}$
\\[-0.15cm]

&&&&&&\\
\hline\hline
\end{tabular}}
\end{table}

\subsubsection{Predicting $(\alpha,\lambda)$ directly from $J(\omega)$}

\textcolor{black}{
A major advantage of the fractional formulation is that $(\alpha,\lambda)$ can be specified directly from microscopic inputs, which are the bath exponent $\chi$ and the asymptotic structure of the dephasing kernel $Q(t)$ (or the structure of spectral density $J(\omega)$). Appendix~\ref{app:alpha-lambda-derivation} provides the full derivations for all three bath classes (sub–Ohmic, Ohmic, super–Ohmic). The resulting predictive rules are summarized in Table~\ref{tab:alpha-lambda-recipe} and are used verbatim in Figure~\ref{fig:no_fitting_sub}. As can be seen from the figure, in the short-time regime ($t<0.2/\omega_c$ the dephasing kernel exhibits the universal Gaussian expansion $Q(t)\approx A_\chi t^{2}$, which fixes the fractional short-time parameters to $\alpha_{\rm pred}=2$ and $\lambda_{\rm pred}=2A_\chi$. While our main fractional master equation is formulated for $0<\alpha<1$, this short-time assignment $\alpha_{\rm pred}=2$ should be understood as an \emph{effective} Caputo order in the scalar coherence equation used solely to match the initial Gaussian curvature; it is not employed as a full fractional Lindblad evolution for all times. Beyond this window, the coherence is governed entirely by the long-time asymptotics of $Q(t)$, and the fractional model transitions to the memory exponent $\alpha_{\rm pred}$ listed in the table. These exponents reproduce the correct non-Markovian decay or saturation behavior over the full long-time interval ($5/\omega_c \le t \le 10^{5}/\omega_c$).}

\textcolor{black}{
The accuracy of these analytic predictions is also illustrated in Figure~\ref{fig:no_fitting_sub}, which compares the parameter-free fractional solutions with exact results for all three bath types. The agreement at early and late times is a direct consequence of the fact that both regimes are controlled by analytic structures of $Q(t)$, and ultimately of $J(\omega)$: the short-time curvature is inherited from the Gaussian prefactor $A_\chi$, while the long-time behavior reflects the power-law (or plateau-anchored) asymptotics of $Q(t)$ determined by the bath exponent $\chi$. As expected, the fractional prediction deviates from the exact dynamics in the intermediate-time interval ($5/\omega_c \lesssim t \lesssim 10^4/\omega_c$). This is precisely the regime where neither the short-time Gaussian form nor the long-time asymptotic expansion of $Q(t)$ is accurate, and where subleading features of the true spectral density $J(\omega)$ play a dominant role. The fact that the discrepancy is confined to this intermediate region suggests that the fractional model may serve as a sensitive diagnostic of missing structure in $J(\omega)$: improving the bath model in this time window would sharpen agreement across all times. Indeed, one may either refine the assumed form of $J(\omega)$ or, conversely, use partial fitting in the intermediate-time range to infer additional information about the underlying spectral density. We explore this perspective in the following subsection.}

\textcolor{black}{
Overall, the comparison demonstrates that the fractional model can be parameterized \emph{predictively and entirely from microscopic inputs}, with the intermediate-time deviations providing useful guidance on how to refine or learn the spectral structure of the environment.}

\begin{figure*}[ht] 
\centering \includegraphics[width=0.95\linewidth]{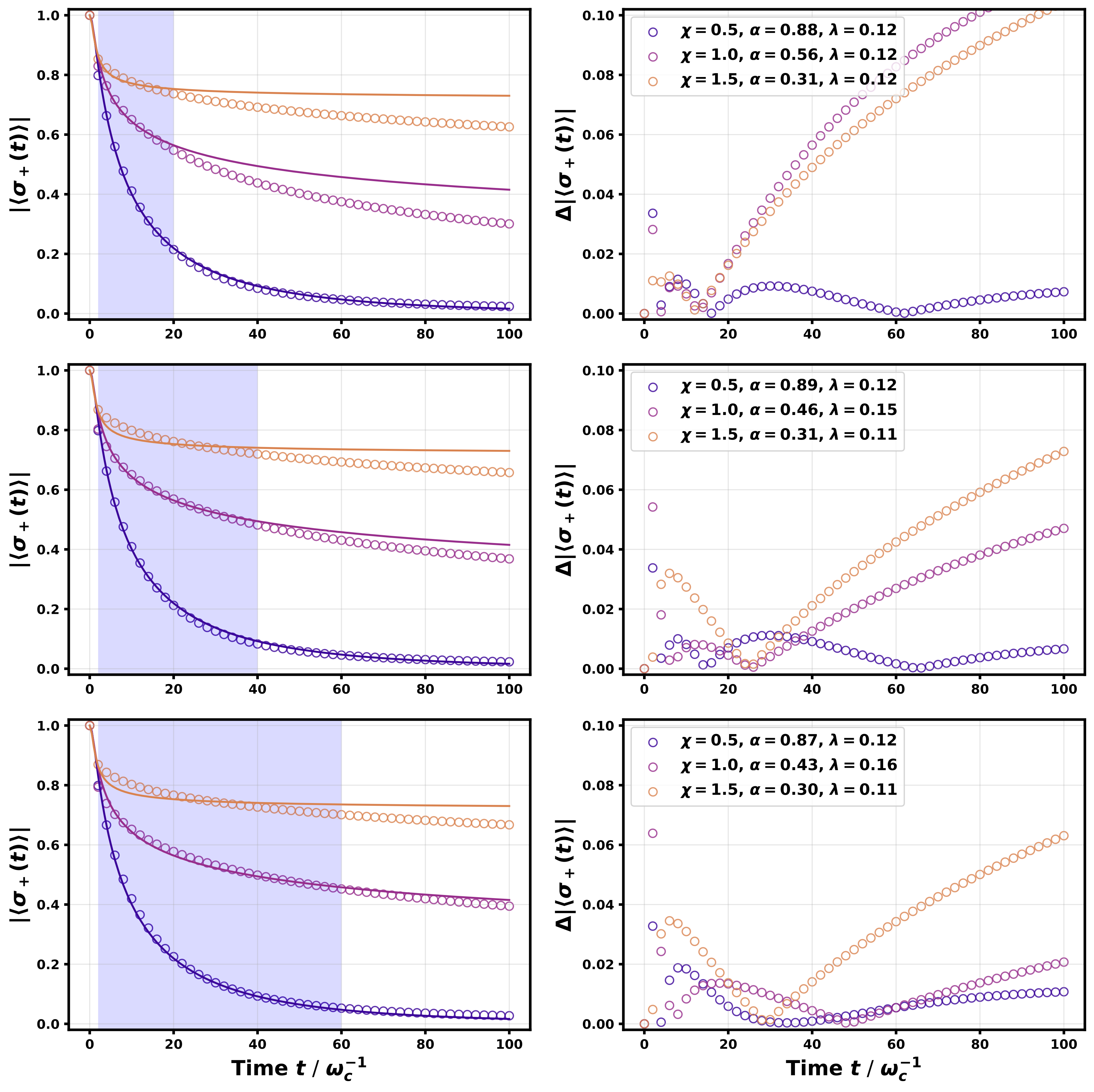} \caption{Asymptotic propagation of the coherence amplitude $|\langle\sigma_+(t)\rangle|$ for the spin--boson model obtained from the single--order fractional propagation and compared with the numerically exact result (solid lines, computed by direct integration of Eq.~\eqref{eq:Q-general}). Results are shown for three spectral exponents, $\chi=0.5$, $1.0$, and $1.5$. The fractional parameters $(\alpha,\lambda)$ are determined by the least--squares fitting to the exact propagation within the indicated intermediate--time windows (blue shades, left panels): $t\!\in\![2/\omega_c,20/\omega_c]$ (top), $t\!\in\![2/\omega_c,40/\omega_c]$ (middle), and $t\!\in\![2/\omega_c,60/\omega_c]$ (bottom). Circle markers denote the fractional results; the shaded regions mark the fitting intervals. \textcolor{black}{The parameters $(\alpha,\lambda)$ shown here are fractional parameters and should not be compared with the Markovian or time-local rates in Figure~\ref{fig:Markovian}. }} \label{fig:frac_general} 
\end{figure*}

\subsubsection{Refinement via least--squares fitting (benchmarking and bath inference)}

\textcolor{black}{
To quantify the accuracy of the single-order fractional model and to extract an effective memory exponent in the intermediate regime where discrepancies emerge, we refine $(\alpha_{\rm pred},\lambda_{\rm pred})$ by least-squares fitting to the exact spin–boson coherence. This refinement is used strictly as a diagnostic and benchmarking tool: it reveals how the fractional parameters summarize bath-induced memory over a specific temporal interval and provides a principled route to \emph{inferring additional structure in $J(\omega)$}. Accordingly, we focus on the interval $5/\omega_c \le t \le 10^{2}/\omega_c$, where the deviation between the predictive fractional model and the exact solution in Figure~\ref{fig:frac_general} is most pronounced. Extending the comparison into the later-time regime ($10^{2}/\omega_c$–$10^{4}/\omega_c$) does not change the qualitative conclusions, as the deviation decreases monotonically once the dynamics enters the asymptotic regime captured by the analytic structure of $Q(t)$.}

Figure~\ref{fig:frac_general} compares the coherence amplitude $|\langle\sigma_+(t)\rangle|$ obtained from the single--order fractional propagation~\eqref{eq:ML} with the numerically exact NZ evolution for representative bath exponents $\chi=0.5$, $1.0$, and $1.5$. For each case, 
The fractional parameters $(\alpha,\lambda)$ are \emph{optimized within the shaded intermediate--time windows} and then the parameters are fixed for the fractional propagation over the entire time domain. 
Concretely, $(\alpha,\lambda)$ are obtained by minimizing the discrete root--mean--square deviation between the fractional coherence $u_\alpha(t;\alpha,\lambda)$ and the NZ result $u_{\mathrm{NZ}}(t)$ over a fitting interval $[t_{\mathrm{start}},t_{\mathrm{end}}]$, cf.\ Appendix~\ref{app:frac-fit}. 
The fitting windows are chosen to be several times the bath correlation time $\tau_B\sim 1/\omega_c$ (so that both the initial curvature and the onset of the long--time tail are resolved) yet shorter than the recurrence time set by the high--frequency cutoff. 

This consistent fitting procedure is applied to the sub--Ohmic, Ohmic, and super--Ohmic examples shown in the top, middle, and bottom panels, respectively. 
As can be seen, the fractional model accurately reproduces the exact coherence for the sub--Ohmic bath (with the deviation $<0.02$ over the entire time domain regardless of the selected time windows for the parameter fitting), demonstrating that a single fractional order can faithfully mimic the memory kernel encoded in $Q(t)$. 
%
Since $Q(t)$ is completely determined by the bath correlation function $C(t)=\langle B(t)B(0)\rangle$ and hence by the spectral density $J(\omega)$, the optimized $(\alpha,\lambda)$ can be viewed as compact, two--parameter summaries of the influence of $C(t)$ over the chosen time window.
For the Ohmic and super--Ohmic baths, the agreement gradually worsens as $\chi$ increases, reflecting the reduced influence of long--tailed correlations. On the other hand, the accuracy in the Ohmic baths can be systematically improved by fitting over a broader time window, which allows the optimized
$(\alpha,\lambda)$ to capture both the short--time curvature and the slower asymptotic decay of the exact solution. Nevertheless, improvements for the super--Ohmic bath are modest compared to the other two baths, and its large residual deviation ($>0.05$) motivates the modified fractional ansatz as described in the following section, which incorporates the finite plateau $u_\infty$ characteristic of super-Ohmic environments.

\textcolor{black}{
More importantly, because the exact coherence spans multiple dynamical regimes---Gaussian at short times, stretched-exponential at intermediate times, and algebraic or plateau-anchored at long times---no single exponent can describe the full evolution. Thus a fitted $\alpha$ should be viewed as an \emph{effective memory exponent} appropriate to the chosen time window, analogous to window-dependent anomalous diffusion exponents in CTRW theory~\cite{klages2008anomalous,METZLER20001,Metzler_2004,Metzler2014}. The variation of fitted $\alpha$ across different windows therefore serves as a quantitative probe of unresolved features in the spectral density $J(\omega)$ and offers a potential route for bath inference. For example, in the sub–Ohmic case, if fitting over an intermediate window yields an effective exponent $\alpha_{\rm fit}$, then matching the stretched-exponential form implies an \emph{effective} spectral exponent $\chi_{\mathrm{eff}} = 1 - \alpha_{\rm fit}$.  Thus $\alpha_{\rm fit} > \alpha_{\rm pred}=1-\chi$ indicates additional low-frequency weight in $J(\omega)$, e.g., a softened cutoff or a secondary power-law component. We present this only as an illustrative example; developing a systematic reconstruction scheme is left for future work.}

\subsubsection{Super--Ohmic baths: fractional ansatz with plateau normalization}

\begin{figure*}[ht]
    \centering
    \includegraphics[width=0.95\linewidth]{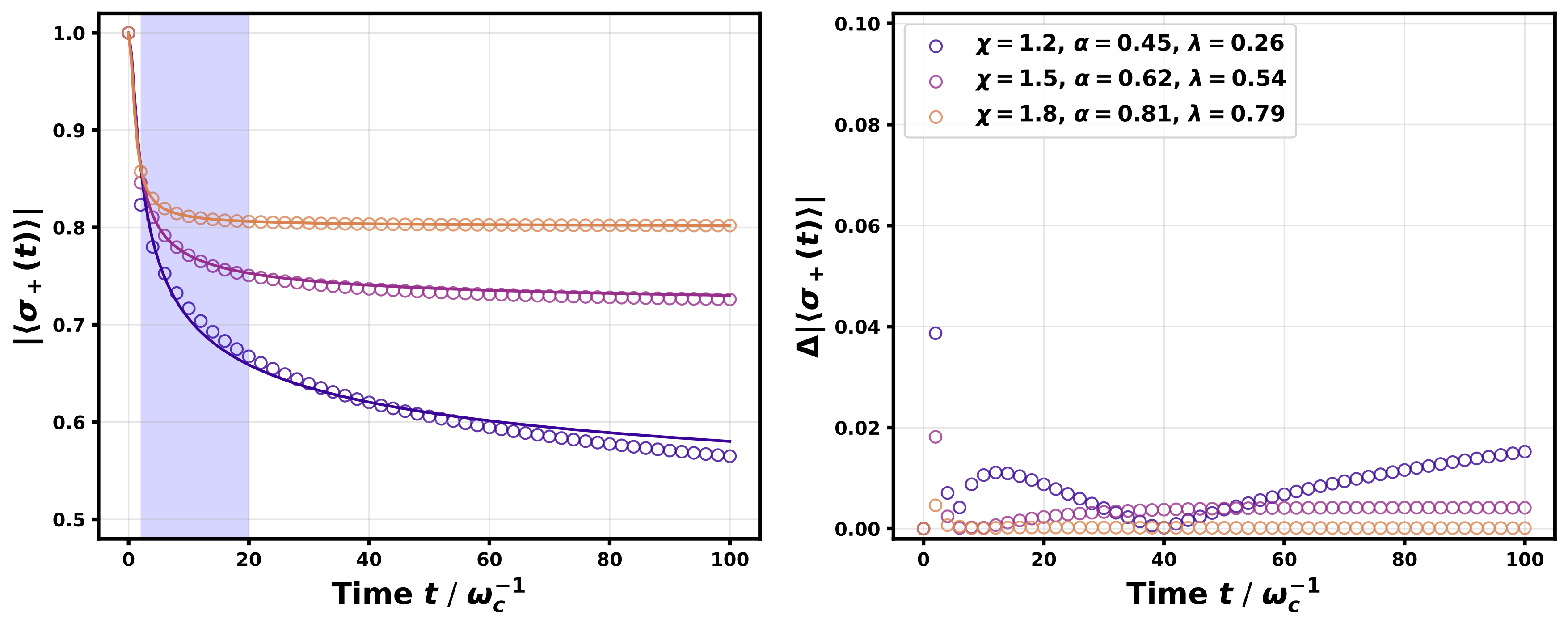}
    \caption{Improved single--order fractional propagation for super--Ohmic baths with spectral exponents $\chi=1.2$, $1.5$, and $1.8$. Fractional propagations are anchored to the finite coherence plateau $u_{\infty}$ (see Eq.~\eqref{eq:ansatz-superohmic}). The fractional orders $(\alpha,\lambda)$ are obtained by least--squares fitting to the numerically exact coherence (solid lines, computed by direct integration of Eq.~\eqref{eq:Q-general}) within the shaded intermediate--time window $t\!\in\![2/\omega_c,20/\omega_c]$, and then the parameters are fixed for the fractional propagation over the entire time domain. Circle markers denote the fractional results, while the right panel shows the absolute deviation from the numerically exact result (solid lines, computed by direct integration of Eq.~\eqref{eq:Q-general}).}
    \label{fig:frac_super}
\end{figure*}

For super--Ohmic baths ($\chi>1$), the coherence does not vanish at long times but saturates to a finite plateau $u_\infty=e^{-Q_\infty}$ determined by the bath reorganization energy. To preserve this limit, the fractional model must include $u_\infty$ explicitly through a convex mixture of the Markovian and fractional channels,
\begin{align}
u(t)&=u_\infty+(1-u_\infty)\,E_{\alpha}(-\lambda t^{\alpha}) \notag \\
\Rightarrow~~
v(t)&=\frac{u(t)-u_\infty}{1-u_\infty}=E_{\alpha}(-\lambda t^{\alpha}). \label{eq:ansatz-superohmic}
\end{align}
This affine form bounds the coherence between $u_\infty$ and~1 and 
\textcolor{black}{renders the initial Gaussian decay and most of the intermediate stretched-exponential behavior negligible, thereby isolating the long-time algebraic regime of the dynamics.}
We therefore fit the \emph{plateau--normalized} curve $v(t)$ within the shaded intermediate--time window to obtain $(\alpha,\lambda)$, and then reconstruct $u(t)$ from Eq.~\eqref{eq:ansatz-superohmic}. The optimized fractional orders are systematically larger than $s-1$, compensating for the overly rapid decay of the fixed--order model and capturing the slower relaxation characteristic of structured baths. As shown in Figure~\ref{fig:frac_super}, this plateau--anchored fractional ansatz accurately reproduces both the intermediate--time curvature and the long--time coherence plateau for all super--Ohmic exponents considered ($\chi=1.2$, $1.5$, and $1.8$).
\textcolor{black}{
Because the asymptotic tail is mapped to a pure power law, the fractional parameters extracted from $v(t)$ in Figure~\ref{fig:frac_super} emphasize the long-time memory exponent and may differ from parameters obtained by fitting the full coherence $u(t)$. For example, for $\chi=1.5$, the fractional order obtained from the plateau-normalized fit ($\alpha = 0.62$) is approximately twice that obtained from the full-coherence fit in Figure~\ref{fig:frac_general} ($\alpha = 0.31$), even when using the same fitting window.}

In practice, when the accessible simulation window lies in the pre--asymptotic regime, the best effective order is slightly larger ($\alpha_{\mathrm{eff}}\!\in(1-\chi,1)$), capturing the stretched--exponential envelope of $u_{\mathrm{NZ}}(t)$ before the Mittag--Leffler power law sets in. A quick, data--driven estimate can be obtained from the local log--log slope of $X(t)=-\ln|u_{\mathrm{NZ}}(t)|$ (for sub-Ohmic and Ohmic baths) or $X(t)=-\ln|v(t)|$ (for super-Ohmic bath) as
\(
\alpha_{\text{loc}}(t_i)=
\big[\ln X(t_{i+1})-\ln X(t_{i-1})\big]/
\big[\ln t_{i+1}-\ln t_{i-1}\big],
\)
with $\alpha$ chosen as the median over the region where the slope is approximately constant. Once $\alpha$ is fixed, $\lambda$ follows from a single point in the same window by solving Eq.~\eqref{eq:ML} or \eqref{eq:ansatz-superohmic}. This \emph{optimization--free} selection rule recovers the expected scaling $\alpha\!\approx\!1-\chi$ for sub--Ohmic baths and provides reliable initial guesses for the least--squares refinement used in Figures~\ref{fig:frac_general} and~\ref{fig:frac_super}.\\

It is worth noting that the single--order fractional Lindbladian in Eq.~\eqref{eq:frac-lindblad} serves as a minimal demonstration because it admits exact analytical benchmarking against the solvable pure--dephasing model. Nevertheless, the underlying construction is fully general: the subordination representation~\eqref{eq:subordination} applies to any bounded generator~$\mathcal{L}$, including dissipative channels that induce population decay, and to composite or multi--qubit systems via tensor--product generators $\mathcal{L}_{\text{tot}}=\sum_i \mathcal{L}_i\otimes\mathbb{I}_{\bar{i}}$. In these cases, the fractional propagation remains CPTP whenever each $\mathcal{L}_i$ is of GKSL form. The subordination framework thus extends beyond pure dephasing and accommodates arbitrary combinations of coherent and incoherent processes. A systematic generalization to distributed--order forms $\int_0^1 \prescript{C}{}{D}_t^{\alpha}\rho\,g(\alpha)\,d\alpha=\mathcal{L}\rho$ is straightforward and provides a flexible route to model strongly structured environments.


\subsection{Unified Interpretation and Quantum Algorithms for Fractional Dynamics}
\label{sec:unified-algorithms}

The fractional framework is able to provide both a unified interpretation of non-Markovian quantum dynamics and a constructive pathway toward their quantum simulation. Numerical benchmarks on the spin–boson model show that a stationary single-order fractional Lindbladian can closely track the exact NZ coherence over broad time windows, with errors at the few-percent level. 
Operationally, the fractional order~$\alpha$ continuously tunes the degree of memory: lowering~$\alpha$ slows relaxation, broadens correlation times, and induces algebraic envelopes, all consistent with the long-tail structure inherited from the bath spectral density~$J(\omega)$.
Within this hierarchy, the Liouville, Lindblad, and fractional master equations correspond respectively to unitary, Markovian, and structured non-Markovian evolution. Fractional differentiation introduces memory directly at the generator level, embedding power-law temporal correlations without violating complete positivity, and providing a continuous interpolation between reversible ($\alpha\!=\!1$) and strongly non-Markovian ($\alpha\!\ll\!1$) limits.

The subordination representation not only clarifies the physical origin of long-memory relaxation but also provides a direct blueprint for quantum simulation. In particular, the Bochner–Phillips subordination representation~\eqref{eq:subordination} replaces explicit memory integration by an average over standard Lindblad semigroups weighted by the inverse-stable density~$f_\alpha(u,t)$. This representation guarantees complete positivity and naturally inspires two complementary quantum-simulation paradigms.

\subparagraph{(i) Polynomial approximation via QSP/QSVT.}
Since the Mittag–Leffler function $E_\alpha(z)$ is entire, the fractional propagator $E_\alpha(\mathcal{L}t^\alpha)$ can be uniformly approximated by bounded polynomials on $\mathrm{spec}(\mathcal{L})$ and implemented through Quantum Signal Processing (QSP) or Quantum Singular Value Transformation (QSVT)~\cite{LowChuang2019,Gilyen2019,Childs2018,Liu2021}, generalizing exponential simulation. 

Let $P_N(z)=\sum_{n=0}^{N} c_n z^n$ with $c_n=1/\Gamma(\alpha n+1)$. For a block-encoded generator
\begin{align}
U_\mathcal{L}= \begin{pmatrix} \mathcal{L}/\lambda_{\max} & \cdot \\[3pt] \cdot & \cdot \end{pmatrix}
\end{align}
(with spectral radius~$\lambda_{\max}$) acting on an extended Hilbert space such that \(\langle 0|U_\mathcal{L}|0\rangle = L/\lambda_{\max}\), $P_N(\mathcal{L}t^\alpha/\lambda_{\max})$ realizes an approximate Mittag–Leffler evolution with deterministic error
\begin{align}
\|E_\alpha(z)-P_N(z)\|
\le |z|^{N+1}/\Gamma[\alpha(N+1)+1].
\end{align}
A degree $N=\mathcal{O}(\sqrt{t^\alpha\lambda_{\max}\log(1/\epsilon)})$ achieves precision~$\epsilon$, giving polylogarithmic scaling in~$1/\epsilon$. This route extends exponential-type algorithms to non-Markovian propagators while retaining the CPTP structure of the underlying Lindblad generator.

\subparagraph{(ii) Subordination sampling (fractional quantum trajectories).}
Alternatively, one may sample operational times~$u$ from~$f_\alpha(u,t)$ and simulate standard Lindblad evolutions~$e^{u\mathcal{L}}\rho(0)$ for each trajectory. Every path corresponds to a physical CPTP evolution of random duration, and observable estimates
\begin{align}
\langle O\rangle_t=\frac{1}{M}\sum_{k=1}^{M}
\mathrm{Tr}[O\,e^{u_k\mathcal{L}}\rho(0)]
\end{align}
converge as~$\mathcal{O}(M^{-1/2})$ with unbiased variance. This defines a \emph{fractional quantum-trajectory method} that emulates long-memory dynamics using only Markovian primitives~\cite{Suess2014} and requires only constant memory per trajectory, since no time-history storage is needed.

In terms of complexity and accuracy, classical fractional solvers scale quadratically in propagation length because they store full history. Both quantum strategies circumvent this cost: the QSP/QSVT approach yields deterministic super-exponential error decay in~$N$ and polylogarithmic gate complexity, whereas the subordination-sampling method scales linearly in trajectory number~$M$ 
(with $M=\mathcal{O}(1/\epsilon^2)$ trajectories sufficient to estimate bounded observables to accuracy $\epsilon$, essentially independent of system size) 
and polylogarithmically in Lindblad simulation cost. Their complementary error behaviors, that is deterministic for QSP and  stochastic~$1/\!\sqrt{M}$ for sampling, make them respectively suited for fault-tolerant and near-term architectures. 
We note that, in principle, the Lindblad evolution $e^{u\mathcal{L}}$ could also be implemented approximately via randomized circuits and engineered noise channels~\cite{dambal2025harnessingintrinsicnoisequantum}, suggesting an interesting direction for combining fractional quantum trajectories with noise engineering on NISQ hardware.

These approaches demonstrate that long-memory quantum processes can be simulated without explicit history storage. By either polynomially approximating the Mittag–Leffler propagator or stochastically averaging over Lindblad evolutions, fractional dynamics emerge as physically consistent CPTP mixtures of Markovian semigroups. This dual theoretical–algorithmic perspective connects open system theory with practical quantum-simulation paradigms, laying the groundwork for scalable studies of non-Markovian dynamics, dissipation, and decoherence within existing hardware toolkits.

While the fault-tolerant strategies discussed above---polynomial approximation via QSP/QSVT and stochastic subordination sampling---provide provably accurate and asymptotically efficient realizations of fractional propagation, recent studies have also pursued near-term hybrid alternatives. 
In particular, Leong \textit{et al.}~\cite{Leong2024} introduced a variational quantum–classical algorithm for solving Caputo time-fractional differential equations, employing overlap history states to capture memory effects on noisy hardware. 
Such variational schemes are complementary to the present framework: they enable proof-of-concept demonstrations of fractional evolution on NISQ devices, whereas the block-encoded and sampling-based approaches developed here are designed for scalable, fault-tolerant architectures with rigorous accuracy and resource guarantees. 


\section{Conclusion and Outlook}

Fractional calculus provides a rigorous and physically consistent extension of quantum master equations for simulating non-Markovian dynamics on both classical and quantum hardware. It embeds long-memory effects within a completely positive framework. By deriving the fractional master equation, proving its equivalence to power-law convolution forms, and expressing it through Lindblad subordination, we have unified three cornerstones of open-system theory---Liouville, Lindblad, and structured non-Markovian evolution---within a single operator-theoretic hierarchy.

Beyond the single-system formulation, the probabilistic subordination picture naturally extends to composite generators and multiqubit systems, since the subordination integral in Eq.~\eqref{eq:subordination} acts operatorwise and preserves tensor–product structure. Hence, the assumption of a GKSL generator is a sufficient but not restrictive condition for guaranteeing complete positivity; non-GKSL extensions can also be analyzed provided the underlying semigroup $e^{u\mathcal{L}}$ remains positive and norm-bounded. 

It is also instructive to contrast this subordination hierarchy with the well-known recursion hierarchies of memory-kernel and time-convolutionless (TCL) master equations~\cite{Breuer2016,Chruscinski2017,deVega2017}. In traditional formulations, higher-order kernels arise from perturbative projections of the system–bath Liouvillian, and complete positivity is generally not preserved at finite truncation order. By contrast, the present construction defines a \emph{stochastic-process hierarchy} in which each level corresponds to a CPTP map generated by a Bochner–Phillips subordination of a Lindblad semigroup. The resulting hierarchy organizes non-Markovian dynamics in terms of nested waiting-time distributions rather than nested kernels, providing a physically transparent and CPTP-preserving alternative to conventional kernel recursion schemes.

This unified picture clarifies how fractional differentiation introduces structured memory directly at the generator level, producing CPTP yet non-divisible dynamics that interpolate smoothly between unitary, Markovian, and strongly non-Markovian regimes. It connects the algebraic tails of long-time relaxation with the complete monotonicity of the subordination kernel and explains why fractional dynamics form a structured subclass of memory-kernel quantum master equations. \textcolor{black}{Moreover, by linking the fractional resolvent to NZ kernels, HEOM self-energies, and influence-functional descriptions, the present framework situates fractional dynamics as a coarse-grained, analytically transparent complement to existing microscopic approaches rather than a replacement for them.}

From an algorithmic perspective, the same structure offers a practical route to simulation. The Mittag–Leffler propagator can be approximated deterministically via polynomial quantum signal processing or stochastically through subordination sampling over operational times. Both approaches realize long-memory quantum dynamics using only Markovian primitives, avoiding explicit time-history storage while preserving complete positivity. This dual deterministic–stochastic formulation establishes a bridge between theoretical open-system analysis and scalable quantum algorithms. \textcolor{black}{Such reduced, generator-level models may be particularly valuable when fully microscopic solvers---such as HEOM, QUAPI, or tensor-network path-integral methods---become computationally expensive, yet one still seeks controlled long-time behavior or interpretable memory parameters.}

Although the present analysis focused on the exactly solvable pure-dephasing limit at zero temperature, the fractional framework itself imposes no such restriction. Because the subordination construction remains valid for any GKSL generator, the same approach can incorporate amplitude damping, population relaxation, and multiqubit tensor-product generators while preserving complete positivity, thus apply to models with dissipation, pumping, and non-commuting system–bath couplings, as clarified in Sec.~\ref{subsec:connections}. \textcolor{black}{The pure-dephasing model was chosen solely because it provides an exact analytical reference against which the fractional formulation can be benchmarked without numerical ambiguity. A natural extension is to apply this framework to dissipative models (with energy relaxation) where only numerical references such as HEOM or QUAPI are available, which will be explored in future work.}

Future research will extend these ideas to distributed-order and space–time-fractional models that capture correlated noise and anomalous diffusion, derive rigorous error bounds for quantum implementations of Mittag–Leffler propagators, and develop hybrid quantum–classical workflows that exploit fractional subordination for efficient noise modeling and dissipation control. More broadly, the fractional framework provides a co-designed theoretical language linking operator theory, stochastic processes, and quantum simulation—offering a unified platform for exploring memory, coherence, and irreversibility in complex quantum systems.


\begin{acknowledgments}
This work is supported by a U.S. Department of Energy, Office of Science, Early Career Research Program award in the Basic Energy Sciences, Division of Chemical Sciences, Geosciences, and Biosciences, Computational and Theoretical Chemistry program under FWP 83466. Pacific Northwest National Laboratory (PNNL) is operated by Battelle for the U.S. DOE under contract DE-AC05-76RL01830. 
Y.Z. acknowledges the support from Laboratory Directed Research and Development (LDRD) program of Los Alamos National Laboratory (LANL). LANL is operated by Triad National Security, LLC, for the National Nuclear Security Administration of the U.S. Department of Energy (contract no. 89233218CNA000001). 
\end{acknowledgments}

\section*{Data Availability Statement}

The data that support the findings of this study are available within the article [and its supplementary material].


\appendix

\section{Caputo--Volterra Equivalence and Convolution-Type QME}
\label{app:caputo-volterra}

We collect the basic operators and prove the equivalence between the Caputo fractional master equation and its Volterra (integral) and convolution (differential) forms, keeping careful track of boundary terms. The equivalence between the Caputo, Volterra, and convolution forms follows standard results in fractional calculus~\cite{Podlubny1999,Kilbas2006,Luchko2020} and their applications to open quantum systems~\cite{Tarasov2021}. The complete monotonicity of the Volterra kernel is a direct consequence of Bernstein’s theorem and the classical subordination results of Bochner and Feller~\cite{Bochner1955,Feller1971}.

\subsection{Operators and identities}
For $\mu>0$, the Riemann--Liouville (RL) fractional integral is given by
\begin{align}
(I_t^{\mu} f)(t)=\frac{1}{\Gamma(\mu)}\int_0^t (t-\tau)^{\mu-1} f(\tau)~d\tau,
\end{align}
and for $0<\alpha<1$, the corresponding Caputo derivative is
\begin{align}
\big(\prescript{C}{}{D}_t^\alpha f\big)(t)
&=\frac{1}{\Gamma(1-\alpha)}\int_0^t (t-\tau)^{-\alpha} f'(\tau)~d\tau \notag \\
&= I_t^{1-\alpha} f'(t).
\end{align}
They satisfy, for sufficiently regular function $f$,
\begin{align}
I_t^{\mu} I_t^{\nu}&=I_t^{\mu+\nu}, \notag \\
\prescript{C}{}{D}_t^\alpha \circ I_t^\alpha f &= f, \label{eq:basic_ids} \\
I_t^\alpha \circ \prescript{C}{}{D}_t^\alpha f &= f - f(0). \notag
\end{align}

\subsection{Equivalence of forms for the fractional Lindblad equation}

\begin{proposition}[Caputo--Volterra--Convolution equivalence]
\label{prop:equiv}
Let $0<\alpha<1$, $\rho$ be absolutely continuous, and $\mathcal{L}$ a (bounded) GKSL generator on the working space. The following are equivalent:
\begin{enumerate}
\item \emph{(Caputo)}:
\begin{align}
\prescript{C}{}{D}_t^\alpha \rho(t)=\mathcal{L}\rho(t).    
\end{align}
\item \emph{(Volterra)}:
\begin{align}
\rho(t)=\rho(0)+\int_0^t K_\alpha^{(\mathrm{V})}(t-\tau)~\mathcal{L}\rho(\tau)~d\tau,    
\end{align}
with $K_\alpha^{(\mathrm{V})}(t)=\frac{t^{\alpha-1}}{\Gamma(\alpha)}$ for $t>0$.
\item \emph{(Convolution, differential)}:
\begin{align}
\dot{\rho}(t)=\int_0^t k_\alpha(t-\tau)~\mathcal{L}\rho(\tau)~d\tau
+\frac{t^{\alpha-1}}{\Gamma(\alpha)}~\mathcal{L}\rho(0),    
\end{align}
with $k_\alpha(t)=\frac{d}{dt}K_\alpha^{(\mathrm{V})}(t)=\frac{t^{\alpha-2}}{\Gamma(\alpha-1)}$ for $t>0$.
\end{enumerate}
If $\mathcal{L}\rho(0)=0$ (e.g., $\rho(0)$ is a fixed point of $\mathcal{L}$), the boundary term vanishes and the differential form is a \emph{pure} convolution.\\

\noindent \textbf{Proof.}
\emph{(Caputo $\Rightarrow$ Volterra):}
Apply $I_t^\alpha$ to the Caputo equation and use \eqref{eq:basic_ids}:
\(
\rho(t)-\rho(0)=I_t^\alpha[\mathcal{L}\rho](t)=\int_0^t K_\alpha^{(\mathrm{V})}(t-\tau)~\mathcal{L}\rho(\tau)~d\tau.
\)
\emph{(Volterra $\Rightarrow$ Convolution):}
Differentiate the Volterra integral (Leibniz rule for weakly singular kernels) to obtain
\(
\dot{\rho}(t)=(k_\alpha * \mathcal{L}\rho)(t) + K_\alpha^{(\mathrm{V})}(t)~\mathcal{L}\rho(0).
\)
\emph{(Convolution $\Rightarrow$ Caputo):}
Integrate the differential form with $I_t^{1-\alpha}$, use $I_t^{1-\alpha}\dot{\rho}=\prescript{C}{}{D}_t^\alpha \rho$, and $I_t^{1-\alpha}k_\alpha=\delta$ in the distributional sense (since $k_\alpha=\frac{d}{dt}K_\alpha^{(\mathrm{V})}$), yielding $\prescript{C}{}{D}_t^\alpha \rho=\mathcal{L}\rho$.
\qed
\end{proposition}

\vspace{0.1cm}
\begin{remark}[Three kernels and their roles]
\label{rem:kernels}
There are three related kernels:
\begin{itemize}
    \item $\kappa_C(t)=\frac{t^{-\alpha}}{\Gamma(1-\alpha)},$  
    \item $K_\alpha^{(\mathrm{V})}(t)=\frac{t^{\alpha-1}}{\Gamma(\alpha)},$
    \item $k_\alpha(t)=\frac{t^{\alpha-2}}{\Gamma(\alpha-1)}.$
\end{itemize}
Only $K_\alpha^{(\mathrm{V})}$ and $k_\alpha=\frac{d}{dt}K_\alpha^{(\mathrm{V})}$ appear as the memory kernels multiplying $\mathcal{L}\rho$ in the integral and differential QME forms, respectively; $\kappa_C$ multiplies $\dot{\rho}$ inside the Caputo derivative.
Moreover, $K_\alpha^{(\mathrm{V})}$ is completely monotone on $\mathbb{R}_+$, hence is the Laplace transform of a positive measure; this underpins the subordination representation of the fractional semigroup.
\end{remark}

\vspace{0.2cm}
The Caputo derivative used throughout this work is preferable for physical initial--value problems because it acts on the first derivative of~$\rho(t)$, ensuring that the initial condition $\rho(0)$ enters linearly and that $\prescript{C}{}{D}_t^{\alpha}\rho(t)\!\to\!\dot{\rho}(t)$ as $\alpha\!\to\!1$. By contrast, the Riemann--Liouville derivative introduces fractional integrals of~$\rho(t)$ that depend explicitly on its entire prehistory, requiring fractional initial conditions that are not directly measurable. In open system dynamics where the state $\rho(0)$ is known, the Caputo definition yields physically meaningful time evolution and consistent reduction to standard Lindblad dynamics in the limit $\alpha\!\to\!1$. Numerically, both definitions produce similar late--time algebraic behavior, but only the Caputo form preserves the conventional initial--state specification used in quantum master equations.


\section{Fractional Adams--Moulton (Volterra) Discretization in Banach Spaces}
\label{app:adams-moulton}

The numerical scheme presented here follows the fractional Adams–Moulton predictor–corrector formulation for Caputo equations~\cite{Diethelm2002,Podlubny1999,Kilbas2006}, widely used for fractional differential equations.

Specifically, we derive an implicit predictor--corrector (Diethelm--Ford--Freed type) scheme for
\begin{align}
\prescript{C}{}{D}_t^\alpha \rho(t)=\mathcal{L}\rho(t),\qquad 0<\alpha<1,    
\end{align}
with $\rho:[0,T]\to\mathcal{B}_1(\mathcal{H})$ absolutely continuous and $\mathcal{L}$ the GKSL generator of a norm-continuous CPTP semigroup. By Prop.~\ref{prop:equiv} (\emph{Caputo $\Leftrightarrow$ Volterra}),
\begin{align}
\label{eq:volterra_form}
\rho(t)=\rho(0)+\frac{1}{\Gamma(\alpha)}\int_0^t (t-\tau)^{\alpha-1}~\mathcal{L}\rho(\tau)~d\tau.
\end{align}

Let $t_n=n h$ with $h=T/N$, and denote $\rho_n\approx \rho(t_n)$. Using piecewise linear (predictor) and trapezoidal (corrector) quadrature of \eqref{eq:volterra_form} yields the standard Lubich weights:
\begin{align}
b_j = (j+1)^{1-\alpha}-j^{1-\alpha}\quad (j\ge 0),
\end{align}
and composite corrector weights $a_{n-k}$ obtained by integrating exactly against $(t_{n+1}-\tau)^{\alpha-1}$ on each subinterval $[t_k,t_{k+1}]$:
\begin{align}
a_{n-k} = (n-k+1)^{\alpha} - 2(n-k)^{\alpha} + (n-k-1)^{\alpha}
\end{align}
with $1\le k\le n-1$, $a_0 = 1^\alpha-0^\alpha$, and $a_n=1^\alpha$ (endpoint corrections are the usual ones for the fractional trapezoid rule).

Employing the predictor--corrector update,
the \emph{predictor} is
\begin{align}
\rho_{n+1}^{\mathrm{(pred)}} = \rho_0 + \frac{h^{\alpha}}{\Gamma(1+\alpha)} 
\sum_{k=0}^{n} b_{n-k}~\mathcal{L}\rho_k ~,
\end{align}
and the \emph{corrector} is implicit:
\begin{align}
\label{eq:corrector}
\rho_{n+1} = \rho_0 + \frac{h^{\alpha}}{\Gamma(1+\alpha)} 
\left[ \sum_{k=0}^{n} a_{n-k}~\mathcal{L}\rho_k + a_{-1}~\mathcal{L}\rho_{n+1}^{\mathrm{(pred)}} \right]
\end{align}
with $a_{-1}=1$. Equivalently,
\begin{align}
\label{eq:linear-solve}
\left(\mathbb{I}-\frac{h^{\alpha}}{\Gamma(1+\alpha)}~\mathcal{L}\right)\rho_{n+1} 
= \rho_0 + \frac{h^{\alpha}}{\Gamma(1+\alpha)} \sum_{k=0}^{n} a_{n-k}~\mathcal{L}\rho_k .
\end{align}
In a finite representation, the left factor is step-invariant and can be factorized once (e.g., using LU/Cholesky factorization).

In terms of the convergence and stability, as discussed in~\citenum{Diethelm2002}, if $\mathcal{L}$ is bounded on the discrete space, the scheme \eqref{eq:corrector}--\eqref{eq:linear-solve} is consistent of order $\mathcal{O}(h^{1+\alpha})$ for sufficiently smooth $\rho$ (local error from polynomial replacement under weakly singular kernel). Also, the scheme is A-stable for the scalar test equation $\prescript{C}{}{D}_t^\alpha y=\lambda y$ with $\Re\lambda\le 0$, in the sense induced by the fractional trapezoid rule.
The proof can be done by employing the Laplace-transform tools and discrete Gr\"{o}nwall inequalities for weakly singular Volterra convolutions.

The naive history cost per step is $\mathcal{O}(n)$. Efficient long-time propagation can be achieved using compressed-memory implementations based on
sum-of-exponentials approximations~\cite{Garrappa2015},
\begin{align}
\frac{t^{\alpha-1}}{\Gamma(\alpha)} \approx \sum_{q=1}^{Q} w_q e^{-\xi_q t},
\end{align}
yields $\mathcal{O}(1)$ history updates via $Q$ auxiliary ODEs, for overall $\mathcal{O}(QN)$ time and $\mathcal{O}(Q)$ memory with controllable error.


\section{Subordination for the Fractional Master Equation}
\label{app:subordination}

The subordination representation used in Sec.~\ref{sec:subordination} follows the classical theory of subordinated semigroups and inverse-stable processes~\cite{Bochner1955,Feller1971,BaeumerMeerschaert2001,MeerschaertSikorskii2012}.

We assume $\mathcal{L}$ is the (time-independent) GKSL generator of a norm-continuous CPTP semigroup
$T(u)=e^{u\mathcal{L}}$ on the trace-class $\mathcal{B}_1(\mathcal{H})$ with growth bound $\omega_0\le 0$. $\rho:[0,\infty)\to \mathcal{B}_1(\mathcal{H})$ is absolutely continuous and of exponential order. 
For $0<\alpha<1$, the Laplace transform of the Caputo derivative satisfies
\begin{align}
\mathbb{L}\big\{\prescript{C}{}{D}_t^\alpha \rho(t)\big\}(s)
= s^\alpha \hat{\rho}(s)-s^{\alpha-1}\rho(0),
\end{align}
where $\hat{\rho}(s)$ is the Laplace transform of $\rho(t)$
\begin{align}
\hat{\rho}(s)=\int_0^\infty e^{-st}\rho(t)~dt,~~~~\Re(s) > 0.
\end{align}
Applying this to Eq.~\eqref{eq:FME} and using $\mathbb{L}\{\mathcal{L}\rho\}(s)=\mathcal{L}\hat{\rho}(s)$ (time-independence of the Lindblad operator) gives
\begin{align}
\label{eq:laplace-resolvent}
&\big(s^\alpha\mathbb{I}-\mathcal{L}\big)\hat{\rho}(s)=s^{\alpha-1}\rho(0) \notag \\
~\Rightarrow~
&\hat{\rho}(s)=s^{\alpha-1}\big(s^\alpha\mathbb{I}-\mathcal{L}\big)^{-1}\rho(0).
\end{align}
To represent the resolvent, recall that if $\mathcal{L}$ generates a strongly continuous ($C_0$) semigroup 
$\{T(u)\}_{u\ge0}$ on a Banach space, then for any $\Re(\lambda)>\omega_0$ (with $\omega_0$ being the growth 
bound of the semigroup) the \emph{resolvent identity} reads as a Bochner integral in operator norm
\begin{align}
(\lambda\mathbb{I}-\mathcal{L})^{-1}
=\int_0^\infty e^{-\lambda u}~T(u)~du.
\end{align}
Here, $T(u)\equiv e^{u\mathcal{L}}$ represents the usual Markovian evolution at time $u$ generated by $\mathcal{L}$.
Setting $\lambda=s^\alpha$ with $\Re(s)>0$, we obtain
\begin{align}
\label{eq:resolvent-bochner}
\big(s^\alpha\mathbb{I}-\mathcal{L}\big)^{-1}
=\int_0^\infty e^{-u s^\alpha}~T(u)~du.
\end{align}
Substituting this expression into Eq.~\eqref{eq:laplace-resolvent} yields
\begin{align}
\label{eq:hat-rho-factorized}
\hat{\rho}(s)
=\int_0^\infty \Big[s^{\alpha-1}e^{-u s^\alpha}\Big]~T(u)\rho(0)~du.
\end{align}

\noindent
The prefactor $s^{\alpha-1}e^{-u s^\alpha}$ acts as the Laplace-domain representation of the kernel that connects the physical time $t$ and 
the operational time $u$. 
We now introduce this kernel explicitly by defining $f_\alpha(u,t)$ as the function whose Laplace transform in $t$ satisfies
\begin{align}
\label{eq:levy-kernel-def}
\mathbb{L}\{f_\alpha(u,t)\}(s)=s^{\alpha-1}e^{-u s^\alpha},
\qquad u\ge 0,~\Re(s)>0.
\end{align}
This property follows from Bernstein’s theorem on completely monotone functions~\cite{Schilling2012Bernstein}. The function $f_\alpha(u,t)$ is known as the \emph{inverse-stable} (or hitting-time) density associated with an 
$\alpha$-stable subordinator.  
It provides a probabilistic bridge between the deterministic operational time $u$ and the physical time $t$, capturing the long-memory and non-Markovian effects introduced by fractional differentiation.
This kernel possesses several key properties that follow directly from Eq.~\eqref{eq:levy-kernel-def}:

\begin{enumerate}
    \item \textbf{Positivity:} 
    $f_\alpha(u,t)\ge0$ for all $u,t\ge0$, as $s^{\alpha-1}e^{-u s^\alpha}$ is a completely monotone function 
    of $s$.

    \item \textbf{Normalization:} 
    $\displaystyle\int_0^\infty f_\alpha(u,t)~du=1$ for each $t\ge0$.  
    Indeed,
    \begin{align}
    \mathbb{L}~\left\{\int_0^\infty f_\alpha(u,t)~du\right\}(s)
    &=\int_0^\infty s^{\alpha-1}e^{-u s^\alpha}~du \notag \\
    &=\frac{s^{\alpha-1}}{s^\alpha} \notag \\
    &=\frac{1}{s},
    \end{align}
    and the inverse Laplace transform of $1/s$ is unity.
\end{enumerate}

Finally, the fractional evolution can be written in the \emph{subordination form}
\begin{align}
\rho_\alpha(t) =\int_0^\infty f_\alpha(u,t)~T(u)\rho(0)~du,    
\end{align}
which expresses the fractional dynamics as a superposition of standard semigroup evolutions 
$T(u)\rho(0)$ weighted by the random-time density $f_\alpha(u,t)$.  
This representation provides both a rigorous analytical foundation and a clear probabilistic interpretation 
for fractional master equations. 
In particular, taking the inverse Laplace transform of \eqref{eq:hat-rho-factorized} in $t$ (and using Fubini/Tonelli to exchange the Bochner and scalar integrals, justified by positivity and exponential bounds) yields Eq.~\eqref{eq:subordination}, which is the desired subordination representation, representing a convex mixture of Lindblad semigroups and remains CPTP, in agreement with the criteria for completely positive dynamical maps discussed in~\cite{Hall2014,Rivas2014}.
\qed


\section{Predictive Determination of $(\alpha,\lambda)$ from the Spectral Density}
\label{app:alpha-lambda-derivation}

\textcolor{black}{
This appendix provides the derivations underlying the predictive rules for $(\alpha_{\rm pred},\lambda_{\rm pred})$ summarized in Table~\ref{tab:alpha-lambda-recipe}. Our goal is to extract the fractional order $\alpha$ and amplitude $\lambda$ directly from microscopic inputs, which are the short- and long-time asymptotic structures of the dephasing kernel $Q(t)$ induced by the spectral density $J(\omega)$.}

\textcolor{black}{
We begin with the exact coherence $u(t)=e^{-Q(t)}$ of the pure–dephasing spin–boson model and compare it to its fractional analogue 
\begin{align}
u_\alpha(t)=E_\alpha(-\lambda t^\alpha),
\end{align}
where $E_\alpha$ denotes the Mittag--Leffler function. A crucial feature of $E_\alpha$ is that it exhibits \emph{two distinct asymptotic regimes}, which must be used appropriately depending on the bath
structure:
\begin{itemize}
\item[(i)] \textbf{Small-argument / moderate-time expansion}
      ($\lambda t^\alpha \ll 1$):
\begin{align}
E_\alpha(-\lambda t^\alpha)
  &= 1 - \frac{\lambda}{\Gamma(1+\alpha)}t^\alpha + \mathcal{O}(t^{2\alpha}) \notag \\
  &\approx
  \exp\Big[-\frac{\lambda}{\Gamma(1+\alpha)} t^\alpha\Big].
\label{eq:ML-small}
\end{align}
This behavior mimics a \emph{stretched exponential} and is the form relevant for matching sub–Ohmic baths, whose exact coherence also decays as a stretched exponential.
\item[(ii)] \textbf{Large-argument / long-time asymptotic}       ($\lambda t^\alpha \gg 1$):
\begin{align}
E_\alpha(-\lambda t^\alpha)
  \sim \frac{1}{\lambda\,\Gamma(1-\alpha)}\, t^{-\alpha}~~~~ (t\rightarrow\infty).
\label{eq:ML-large}
\end{align}
This algebraic decay is the correct asymptotic structure for matching Ohmic baths (which yield power-law decoherence) and super–Ohmic baths (where the coherence relaxes to its plateau with a power-law correction).
\end{itemize}
The correct identification of $(\alpha,\lambda)$ therefore depends on whether the exact coherence $u(t)$ (determined by the long-time form of $Q(t)$) is stretched exponential (sub–Ohmic) or algebraic (Ohmic or super–Ohmic). In the sections below, we match the appropriate Mittag--Leffler regime to the asymptotic expansions of $Q(t)$ that follow from the low-frequency structure of the spectral density shown in Eq.~\eqref{eq:J-omega}.
}

\textcolor{black}{
\paragraph{Universal short-time regime}
The short-time expansion of $Q(t)$ is universal for all bath types. Expanding the cosine in Eq.~(\ref{eq:Q-general}) gives
\begin{align}
Q(t)= A_\chi\, t^2 + \mathcal{O}(t^4),
~~\text{with}~~A_\chi = \frac{\eta\,\Gamma(\chi+1)\,\omega_c^2}{\pi}.
\end{align}
Thus the exact coherence behaves as
\begin{align}
u(t) &= \exp(-A_\chi\, t^2 + \mathcal{O}(t^4)) \notag \\
&= 1 - A_\chi t^2 + \mathcal{O}(t^4).    
\end{align}
The Mittag–Leffler expansion, Eq.~\eqref{eq:ML-small}, matches the curvature only when 
\begin{align}
    \alpha_{\rm pred}=2,~~\lambda_{\rm pred}=2A_\chi~.
\end{align}
These relations provide the analytic short-time assignments used in Figure~\ref{fig:no_fitting_sub} and in the main text. They require only the microscopic coefficient $A_\chi$, and involve no fitting.}

\textcolor{black}{
\paragraph{Sub–Ohmic baths ($0<\chi<1$)}
For $J(\omega)\propto \omega^\chi$ with $0<\chi<1$, the long-time dephasing is dominated by low-frequency modes. The asymptotic form is
\begin{align}
    Q(t) \sim C_\chi\, t^{1-\chi}~~~~ (t\to\infty),
\end{align}
with $C_\chi$ determined by the infrared part of the spectral density. The corresponding coherence decays as
\begin{align}
u(t)&\sim e^{-C_\chi t^{1-\chi}} \notag \\
&\sim 1 -C_\chi t^{1-\chi} + \mathcal{O}(t^{2-2\chi}).    
\end{align}
Matching the leading power-law behavior of $E_\alpha(-\lambda t^\alpha)$ in Eq.~\eqref{eq:ML-small} yields
\begin{align}
    \alpha_{\rm pred}=1-\chi,~~
    \lambda_{\rm pred}= C_\chi\, \Gamma(2-\chi)~.
\end{align}
}

\textcolor{black}{
\paragraph{Ohmic bath ($\chi=1$)}
For the Ohmic case,
\begin{align}
    J(\omega)=\eta\,\omega\,e^{-\omega/\omega_c},
\end{align}
the exact long-time form is
\begin{align}
Q(t)\sim \frac{2\eta}{\pi}\,\ln t + \text{const} ~~  \rightarrow~~
u(t)\sim t^{-2\eta/\pi}.
\end{align}
Matching the Mittag–Leffler asymptotic in Eq.~\eqref{eq:ML-large} gives
\begin{align}
    \alpha_{\rm pred}=\frac{2\eta}{\pi},~~\lambda_{\rm pred} = \frac{1}{\Gamma(1-\alpha_{\rm pred})}~.
\end{align}
}

\textcolor{black}{
\paragraph{Super–Ohmic baths ($\chi>1$)}
For $\chi>1$, the dephasing saturates at a finite value:
\begin{align}
Q_\infty = \frac{2\eta}{\pi}\,\Gamma(\chi-1),~~
u_\infty = e^{-Q_\infty}.    
\end{align}
The asymptotic correction takes the form
\begin{align}
Q(t) = Q_\infty - D_\chi\, t^{-(\chi-1)} + \cdots,
\end{align}
so that
\begin{align}
u(t)-u_\infty 
\sim u_\infty\, D_\chi\, t^{-(\chi-1)}.
\end{align}
Define the plateau-normalized coherence $v(t)$ as Eq.~\eqref{eq:ansatz-superohmic}, then
\begin{align}
v(t)\sim \frac{u_\infty}{1-u_\infty}\,D_\chi\, t^{-(\chi-1)}.    
\end{align}
Matching the Mittag–Leffler form Eq.~\eqref{eq:ML-large} to above $v(t)$ expression yields
\begin{align}
    \alpha_{\rm pred}= \chi-1,~~\lambda_{\rm pred}
= \frac{e^{Q_\infty}-1}{D_\chi\,\Gamma(2-\chi)}.
\end{align}
}

\textcolor{black}{
\paragraph{Summary}
The microscopic short- and long-time structures of $Q(t)$ fully determine the fractional parameters $(\alpha_{\rm pred},\lambda_{\rm pred})$. Short-time matching always gives $(\alpha,\lambda)=(2,2A_\chi)$, while the long-time exponents and amplitudes follow directly from the infrared behavior of the bath through the power $\chi$. The results for all three regimes match those listed in Table~\ref{tab:alpha-lambda-recipe}, and are precisely the values used in the parameter-free comparisons of Sec.~III.B and Figure~\ref{fig:no_fitting_sub}.}


\section{Optimization of fractional parameters}
\label{app:frac-fit}

For each bath exponent $\chi$, the single--order fractional model is specified by
the propagator
\begin{align}
u_\alpha(t;\alpha,\lambda)=E_\alpha\!\big(-\lambda t^\alpha\big),
\end{align}
which is compared to the numerically exact NZ coherence $u_{\mathrm{NZ}}(t)=\exp[-Q(t)]$ obtained from the structured spectral density $J(\omega)$. The parameters $(\alpha,\lambda)$ are determined by a least--squares fit over a
finite time window $[t_{\mathrm{start}},t_{\mathrm{end}}]$:
\begin{align}
\mathrm{RMSE}^2(\alpha,\lambda)
=\frac{1}{N_{\mathrm{fit}}}\sum_{k:\,t_k\in[t_{\mathrm{start}},t_{\mathrm{end}}]}
\Big|u_\alpha(t_k;\alpha,\lambda)-u_{\mathrm{NZ}}(t_k)\Big|^2,
\end{align}
with $(\alpha,\lambda)$ chosen to minimize $\mathrm{RMSE}^2(\alpha,\lambda)$. In practice we use a coarse grid search followed by a local refinement (Nelder--Mead) to identify the optimal parameters.

The fitting window is guided by the bath correlation time. In our dimensionless units we set $\omega_c = 1$, so the bath correlation time is of order $\tau_B \sim 1/\omega_c = 1$.  More quantitatively, we estimate $\tau_B$ from the decay of the bath correlation function $C(t) = \langle B(t) B(0)\rangle$ (or equivalently from $Q''(t)$) as the time at which $|C(t)|$ has dropped below $e^{-1}|C(0)|$.  The fitting windows are then chosen as $t_{\mathrm{start}} \approx 2\tau_B$ and $t_{\mathrm{end}} \in [20\tau_B, 60\tau_B]$, which in practice corresponds to $[2/\omega_c, 20/\omega_c]$–$[2/\omega_c, 60/\omega_c]$ in Figure~\ref{fig:frac_general}. This ensures that the optimized $(\alpha,\lambda)$ are extracted from a regime where the bath has largely decorrelated but before finite-cutoff recurrences
become important.

Since $Q(t)$ is obtained by a double time integral over the bath correlation function $C(t)$, the optimized pair $(\alpha,\lambda)$ should be regarded as an effective two--parameter representation of the influence of $C(t)$ over the chosen fitting window.  In this sense, $\alpha$ controls the effective long--time decay law of the coherence, while $\lambda$ sets the overall decoherence timescale.  For general structured spectral densities there is no closed-form analytic relation between $C(t)$ and $(\alpha,\lambda)$; the fractional parameters summarize the memory characteristics encoded in $Q(t)$ rather than reproduce them exactly.

\bibliography{aipsamp}

@article{Chen:2025to,
        author = {Chen, Xinxian and Franco, Ignacio},
        doi = {10.1063/5.0278591},
        isbn = {0021-9606},
        journal = {J. Chem. Phys.},
        journal1 = {J. Chem. Phys.},
        month = {12/10/2025},
        number = {10},
        pages = {104109},
        title = {Tree tensor network hierarchical equations of motion based on time-dependent variational principle for efficient open quantum dynamics in structured thermal environments},
        url = {https://doi.org/10.1063/5.0278591},
        volume = {163},
        year = {2025},
        year1 = {2025/09/08}}

@article{Yan:2021tg,
        author = {Yan, Yaming and Xu, Meng and Li, Tianchu and Shi, Qiang},
        doi = {10.1063/5.0050720},
        isbn = {0021-9606},
        journal = {J. Chem. Phys.},
        journal1 = {J. Chem. Phys.},
        month = {12/10/2025},
        number = {19},
        pages = {194104},
        title = {Efficient propagation of the hierarchical equations of motion using the Tucker and hierarchical Tucker tensors},
        url = {https://doi.org/10.1063/5.0050720},
        volume = {154},
        year = {2021},
        year1 = {2021/05/17}}

@article{dambal2025harnessingintrinsicnoisequantum,
      title={Harnessing Intrinsic Noise for Quantum Simulation of Open Quantum Systems}, 
      author={Sameer Dambal and Akira Sone and Yu Zhang},
      year={2025},
      pages={arXiv:2510.21075},
      journal={arXiv Preprint} 
}

@article{Leggett1987,
  title = {Dynamics of the dissipative two-state system},
  author = {Leggett, A. J. and Chakravarty, S. and Dorsey, A. T. and Fisher, Matthew P. A. and Garg, Anupam and Zwerger, W.},
  journal = {Rev. Mod. Phys.},
  volume = {59},
  issue = {1},
  pages = {1--85},
  numpages = {0},
  year = {1987},
  month = {Jan},
  publisher = {American Physical Society},
  doi = {10.1103/RevModPhys.59.1}
}

@article{Gorini1976,
  author  = {Gorini, V. and Kossakowski, A. and Sudarshan, E. C. G.},
  title   = {Completely positive dynamical semigroups of $N$-level systems},
  journal = {J. Math. Phys.},
  volume  = {17},
  pages   = {821},
  year    = {1976},
  doi     = {10.1063/1.522979}
}

@article{Lindblad1976,
  author  = {Lindblad, G.},
  title   = {On the generators of quantum dynamical semigroups},
  journal = {Commun. Math. Phys.},
  volume  = {48},
  pages   = {119--130},
  year    = {1976},
  doi     = {10.1007/BF01608499}
}

@book{vonNeumann1929,
  title={Mathematical Foundations of Quantum Mechanics},
  author={von Neumann, J.},
  isbn={9780691028934},
  lccn={53010143},
  series={Goldstine Printed Materials},
  url={https://books.google.com/books?id=JLyCo3RO4qUC},
  year={1955},
  publisher={Princeton University Press}
}

@article{Alicki1977,
title = {On the detailed balance condition for non-hamiltonian systems},
journal = {Rep. Math. Phys.},
volume = {10},
number = {2},
pages = {249-258},
year = {1976},
issn = {0034-4877},
doi = {10.1016/0034-4877(76)90046-X},
author = {Robert Alicki},
}

@book{Hilfer2000,
  title={Applications Of Fractional Calculus In Physics},
  author={Hilfer, R.},
  isbn={9789814496209},
  url={https://books.google.com/books?id=MIXVCgAAQBAJ},
  year={2000},
  publisher={World Scientific Publishing Company}
}

@article{Mainardi2000,
title = {On Mittag-Leffler-type functions in fractional evolution processes},
journal = {J. Comput. Appl. Math.},
volume = {118},
number = {1},
pages = {283-299},
year = {2000},
issn = {0377-0427},
doi = {10.1016/S0377-0427(00)00294-6},
author = {Francesco Mainardi and Rudolf Gorenflo}
}

@book{Breuer2002,
  title={The Theory of Open Quantum Systems},
  author={Breuer, H.P. and Petruccione, F.},
  isbn={9780198520634},
  lccn={2002075713},
  url={https://books.google.com/books?id=0Yx5VzaMYm8C},
  year={2002},
  publisher={Oxford University Press}
}

@book{Bochner1955,
  title={Harmonic Analysis and the Theory of Probability},
  author={Bochner, S.},
  isbn={9780520372535},
  series={California monographs in mathematical sciences},
  url={https://books.google.com/books?id=Akx-EAAAQBAJ},
  year={2022},
  publisher={University of California Press}
}

@book{Nielsen2000,
  title={Quantum Computation and Quantum Information: 10th Anniversary Edition},
  author={Nielsen, M.A. and Chuang, I.L.},
  isbn={9781139495486},
  url={https://books.google.com/books?id=-s4DEy7o-a0C},
  year={2010},
  publisher={Cambridge University Press}
}

@article{Chruscinski2017,
  title={A Brief History of the GKLS Equation},
  author={Dariusz Chruściński and Saverio Pascazio},
  journal={Open Syst. Inf. Dyn.},
  year={2017},
  volume={24},
  pages={1740001},
  doi={10.1142/S1230161217400017}
}

@article{Breuer2016,
  author  = {Breuer, Heinz-Peter and Laine, Elsi-Mari and Piilo, Jyrki and Vacchini, Bassano},
  title   = {Non-Markovian dynamics in open quantum systems},
  journal = {Rev. Mod. Phys.},
  volume  = {88},
  pages   = {021002},
  year    = {2016},
  doi     = {10.1103/RevModPhys.88.021002}
}

@article{Li2018,
  author  = {Li, Li and Hall, Michael J. W. and Wiseman, Howard M.},
  title   = {Concepts of quantum non-Markovianity: a hierarchy},
  journal = {Phys. Rep.},
  volume  = {759},
  pages   = {1--51},
  year    = {2018},
  doi     = {10.1016/j.physrep.2018.07.001}
}

@article{Nakajima1958,
    author = {Nakajima, Sadao},
    title = {On Quantum Theory of Transport Phenomena: Steady Diffusion},
    journal = {Prog. Theor. Phys.},
    volume = {20},
    number = {6},
    pages = {948-959},
    year = {1958},
    month = {12},
    doi = {10.1143/PTP.20.948}
}

@article{Zwanzig1960,
  author  = {Zwanzig, R.},
  title   = {Ensemble method in the theory of irreversibility},
  journal = {J. Chem. Phys.},
  volume  = {33},
  pages   = {1338--1341},
  year    = {1960},
  doi     = {10.1063/1.1731409}
}

@book{Gorenflo2014,
  title={Mittag-Leffler Functions, Related Topics and Applications},
  author={Gorenflo, R. and Kilbas, A.A. and Mainardi, F. and Rogosin, S.},
  isbn={9783662615508},
  series={Springer Monographs in Mathematics},
  url={https://books.google.com/books?id=kZUFEAAAQBAJ},
  year={2020},
  publisher={Springer Berlin Heidelberg}
}

@Inbook{Tarasov2021,
author="Tarasov, Vasily E.",
title="Fractional Dynamics of Open Quantum Systems",
bookTitle="Fractional Dynamics: Applications of Fractional Calculus to Dynamics of Particles, Fields and Media",
year="2010",
publisher="Springer Berlin Heidelberg",
address="Berlin, Heidelberg",
pages="467--490",
isbn="978-3-642-14003-7",
doi="10.1007/978-3-642-14003-7_20",
url="https://doi.org/10.1007/978-3-642-14003-7_20"
}

@book{Podlubny1999,
  title={Fractional Differential Equations: An Introduction to Fractional Derivatives, Fractional Differential Equations, to Methods of Their Solution and Some of Their Applications},
  author={Podlubny, I.},
  isbn={9780080531984},
  series={Mathematics in Science and Engineering},
  url={https://books.google.com/books?id=K5FdXohLto0C},
  year={1998},
  publisher={Academic Press}
}

@article{deVega2017,
  author   = {de Vega, Inés and Alonso, Daniel},
  title    = {Dynamics of non-Markovian open quantum systems},
  journal  = {Rev. Mod. Phys.},
  volume   = {89},
  pages    = {015001},
  year     = {2017},
  doi      = {10.1103/RevModPhys.89.015001}
}

@book{Feller1971,
  title={An Introduction to Probability Theory and Its Applications, Volume 2},
  author={Feller, W.},
  isbn={9780471257097},
  lccn={68011708},
  series={Wiley Series in Probability and Statistics},
  url={https://books.google.com/books?id=rxadEAAAQBAJ},
  year={1991},
  publisher={Wiley}
}

@article{Rivas2014,
  author   = {Rivas, \'{A}ngel and Huelga, Susana F. and Plenio, Martin B.},
  title    = {Quantum non-Markovianity: characterization, quantification and detection},
  journal  = {Rep. Prog. Phys.},
  volume   = {77},
  pages    = {094001},
  year     = {2014},
  doi      = {10.1088/0034-4885/77/9/094001}
}

@book{Kilbas2006,
  title={Theory And Applications of Fractional Differential Equations},
  author={Kilbas, A.A.A. and Srivastava, H.M. and Trujillo, J.J.},
  isbn={9780444518323},
  lccn={05044764},
  series={North-Holland Mathematics Studies},
  url={https://books.google.com/books?id=LhkO83ZioQkC},
  year={2006},
  publisher={Elsevier Science \& Tech}
}

@article{Metzler2014,
author ="Metzler, Ralf and Jeon, Jae-Hyung and Cherstvy, Andrey G. and Barkai, Eli",
title  ="Anomalous diffusion models and their properties: non-stationarity{,} non-ergodicity{,} and ageing at the centenary of single particle tracking",
journal  ="Phys. Chem. Chem. Phys.",
year  ="2014",
volume  ="16",
issue  ="44",
pages  ="24128-24164",
publisher  ="The Royal Society of Chemistry",
doi  ="10.1039/C4CP03465A"
}

@article{Hall2014,
  title = {Canonical form of master equations and characterization of non-Markovianity},
  author = {Hall, Michael J. W. and Cresser, James D. and Li, Li and Andersson, Erika},
  journal = {Phys. Rev. A},
  volume = {89},
  issue = {4},
  pages = {042120},
  numpages = {11},
  year = {2014},
  month = {Apr},
  publisher = {American Physical Society},
  doi = {10.1103/PhysRevA.89.042120},
  url = {https://link.aps.org/doi/10.1103/PhysRevA.89.042120}
}

@book{Weiss2012,
  title={Quantum Dissipative Systems},
  author={Weiss, U.},
  isbn={9789812791627},
  lccn={2008274341},
  series={Series in modern condensed matter physics},
  url={https://books.google.com/books?id=iV9pDQAAQBAJ},
  year={2008},
  publisher={World Scientific}
}

@article{LowChuang2019,
  author    = {Guang Hao Low and Isaac L. Chuang},
  title     = {Hamiltonian Simulation with Optimal Sample Complexity},
  journal   = {Quantum},
  volume    = {3},
  pages     = {163},
  year      = {2019},
  doi       = {10.22331/q-2019-07-12-163}
}

@inproceedings{Gilyen2019,
  author    = {Andr{\'a}s Gily{\'e}n and Yuan Su and Guang Hao Low and Nathan Wiebe},
  title     = {Quantum Singular Value Transformation and Beyond: Exponential Improvements for Quantum Matrix Arithmetics},
  booktitle = {Proceedings of the 51st Annual ACM SIGACT Symposium on Theory of Computing (STOC 2019)},
  pages     = {193--204},
  year      = {2019},
  doi       = {10.1145/3313276.3316366}
}

@article{Childs2018,
	author = {Andrew M. Childs and Dmitri Maslov and Yunseong Nam and Neil J. Ross and Yuan Su},
	doi = {10.1073/pnas.1801723115},
	journal = {Proc. Natl. Acad. Sci. U.S.A.},
	number = {38},
	pages = {9456--9461},
	title = {Toward the First Quantum Simulation with Quantum Speedup},
	volume = {115},
	year = {2018}}

@article{Murch2013,
  author    = {K. W. Murch and S. J. Weber and K. M. Beck and E. Ginossar and I. Siddiqi},
  title     = {Observing single quantum trajectories of a superconducting quantum bit},
  journal   = {Nature},
  volume    = {502},
  pages     = {211--214},
  year      = {2013},
  doi       = {10.1038/nature12539}
}

@article{Kjaergaard2020,
  author    = {M. Kjaergaard and M. E. Schwartz and J. Braum{\"u}ller and P. Krantz and J. I.-J. Wang and S. Gustavsson and W. D. Oliver},
  title     = {Superconducting qubits: Current state of play},
  journal   = {Annu. Rev. Condens. Matter Phys.},
  volume    = {11},
  pages     = {369--395},
  year      = {2020},
  doi       = {10.1146/annurev-conmatphys-031119-050605}
}

@article{PhysRevLett.121.060401,
  title   = {Controllable Non-Markovianity for a Spin Qubit in Diamond},
  author  = {Haase, J. F. and Vetter, P. J. and Unden, T. and Smirne, A. and Rosskopf, J. and Naydenov, B. and Stacey, A. and Jelezko, F. and Plenio, M. B. and Huelga, S. F.},
  journal = {Phys. Rev. Lett.},
  volume  = {121},
  number  = {6},
  pages   = {060401},
  year    = {2018},
  doi     = {10.1103/PhysRevLett.121.060401}
}

@article{Clos2016,
  author    = {G. Clos and D. Porras and U. Warring and T. Schaetz},
  title     = {Time-Resolved Observation of Thermalization in an Isolated Quantum System},
  journal   = {Phys. Rev. Lett.},
  volume    = {117},
  number    = {17},
  pages     = {170401},
  year      = {2016},
  doi       = {10.1103/PhysRevLett.117.170401}
}

@article{Vacchini2011,
  author    = {Vacchini, Bassano and Smirne, Andrea and Laine, Elsi-Mari and Piilo, Jyrki and Breuer, Heinz-Peter},
  title     = {Markovianity and non-Markovianity in quantum and classical systems},
  journal   = {New J. Phys.},
  volume    = {13},
  number    = {9},
  pages     = {093004},
  year      = {2011},
  doi       = {10.1088/1367-2630/13/9/093004}
}

@article{Dominy2015,
  title        = {Beyond Complete Positivity},
  author       = {Jason M. Dominy and Daniel A. Lidar},
  year         = {2015},
  journal      = {arXiv Preprint},
  pages        = {arXiv:1503.05342}
}

@article{PhysRevA.87.012127,
  title     = {Non-Markovian probes in ultracold gases},
  author    = {Haikka, P. and McEndoo, S. and Maniscalco, S.},
  journal   = {Phys. Rev. A},
  volume    = {87},
  number    = {1},
  pages     = {012127},
  year      = {2013},
  doi       = {10.1103/PhysRevA.87.012127}
}

@article{Luchko2020,
  author    = {Y. Luchko},
  title     = {Fractional derivatives and the fundamental theorem of fractional calculus},
  journal   = {Fract. Calc. Appl. Anal.},
  volume    = {23},
  pages     = {939--966},
  year      = {2020},
  doi       = {10.1515/fca-2020-0049}
}

@article{Diethelm2002,
  author    = {K. Diethelm and N. J. Ford and A. D. Freed},
  title     = {A predictor-corrector approach for the numerical solution of fractional differential equations},
  journal   = {Nonlinear Dyn.},
  volume    = {29},
  pages     = {3--22},
  year      = {2002},
  doi       = {10.1023/A:1016592219341}
}

@book{Schilling2012Bernstein,
  author    = {R. L. Schilling and R. Song and Z. Vondracek},
  title     = {Bernstein Functions: Theory and Applications},
  series    = {de Gruyter Studies in Mathematics},
  edition   = {2},
  publisher = {de Gruyter},
  year      = {2012},
  doi       = {10.1515/9783110269338}
}

@book{MeerschaertSikorskii2012,
  author    = {M. M. Meerschaert and A. Sikorskii},
  title     = {Stochastic Models for Fractional Calculus},
  series    = {de Gruyter Studies in Mathematics},
  publisher = {de Gruyter},
  year      = {2012},
  doi       = {10.1515/9783110258165}
}

@article{BaeumerMeerschaert2001,
  author  = {Boris Baeumer and Mark M. Meerschaert},
  title   = {Stochastic solutions for fractional Cauchy problems},
  journal = {Fractional Calculus and Applied Analysis},
  volume  = {4},
  pages   = {481--500},
  year    = {2001},
  month   = {Mar}
}

@article{Grifoni1998,
  author  = {Milena Grifoni and Peter H{\"a}nggi},
  title   = {Driven quantum tunneling},
  journal = {Physics Reports},
  volume  = {304},
  pages   = {229--354},
  year    = {1998},
  doi     = {10.1016/S0370-1573(98)00022-2}
}

@article{Liu2021,
author = {Jin-Peng Liu  and Herman Øie Kolden  and Hari K. Krovi  and Nuno F. Loureiro  and Konstantina Trivisa  and Andrew M. Childs },
title = {Efficient quantum algorithm for dissipative nonlinear differential equations},
journal = {Proc. Natl. Acad. Sci. U.S.A.},
volume = {118},
number = {35},
pages = {e2026805118},
year = {2021},
doi = {10.1073/pnas.2026805118}
}

@article{Suess2014,
  author  = {Daniel Suess and Andreas Eisfeld and Walter Strunz},
  title   = {Hierarchy of stochastic pure states for open quantum system dynamics},
  journal = {Phys. Rev. Lett.},
  volume  = {113},
  number  = {15},
  pages   = {150403},
  year    = {2014},
  doi     = {10.1103/PhysRevLett.113.150403}
}

@article{Leong2024,
  author  = {Fong Yew Leong and Dax Enshan Koh and Jian Feng Kong and Siong Thye Goh and Jun Yong Khoo and Wei-Bin Ewe and Hongying Li and Jayne Thompson and Dario Poletti},
  title   = {Solving fractional differential equations on a quantum computer: A variational approach},
  journal = {AVS Quantum Sci.},
  volume  = {6},
  number  = {3},
  pages   = {033802},
  year    = {2024},
  doi     = {10.1116/5.0202971}
}

@article{Loss1998,
  title = {Quantum computation with quantum dots},
  author = {Loss, Daniel and DiVincenzo, David P.},
  journal = {Phys. Rev. A},
  volume = {57},
  issue = {1},
  pages = {120--126},
  numpages = {0},
  year = {1998},
  month = {Jan},
  publisher = {American Physical Society},
  doi = {10.1103/PhysRevA.57.120}
}

@article{Garrappa2015,
  author  = {Roberto Garrappa},
  title   = {Numerical solution of fractional differential equations: A survey and a software tutorial},
  journal = {Mathematics},
  volume  = {3},
  number  = {2},
  pages   = {337--374},
  year    = {2015},
  doi     = {10.3390/math3020337}
}

@article{lopez2025generalisedfractionalrabiproblem, 
title={Generalised fractional Rabi problem}, 
author={Alexander Lopez and Sébastien Fumeron and Malte Henkel and Trifce Sandev and Esther D. Gutiérrez}, 
year={2025}, 
pages={arXiv:2510.08167}, 
Journal={arXiv Preprint}
}

@article{Ang2024MemoryInducedWeakDissipation,
    author = {Kee, Chun Yun and Ang, L. K.},
    title = {Memory-induced weak dissipation in fractional-time-derivative quantum Lindblad-based model},
    journal = {APL Quantum},
    volume = {1},
    number = {1},
    pages = {016112},
    year = {2024},
    month = {03},
    doi = {10.1063/5.0194452}
}

@article{Sandev2018FCAA,
  author  = {T. Sandev and R. Metzler and A. V. Chechkin},
  title   = {From continuous time random walks to the generalized diffusion equation},
  journal = {Fract. Calc. Appl. Anal.},
  volume  = {21},
  pages   = {10--28},
  year    = {2018},
  doi     = {10.1515/fca-2018-0002}
}

@article{PhysRevLett.112.120404,
  title = {Degree of Non-Markovianity of Quantum Evolution},
  author = {Chru\ifmmode \acute{s}\else \'{s}\fi{}ci\ifmmode \acute{n}\else \'{n}\fi{}ski, Dariusz and Maniscalco, Sabrina},
  journal = {Phys. Rev. Lett.},
  volume = {112},
  issue = {12},
  pages = {120404},
  numpages = {5},
  year = {2014},
  month = {Mar},
  publisher = {American Physical Society},
  doi = {10.1103/PhysRevLett.112.120404},
  url = {https://link.aps.org/doi/10.1103/PhysRevLett.112.120404}
}

@article{Sandev_2019,
doi = {10.1088/1751-8121/aaefa3},
url = {https://doi.org/10.1088/1751-8121/aaefa3},
year = {2018},
month = {nov},
publisher = {IOP Publishing},
volume = {52},
number = {1},
pages = {015201},
author = {Sandev, Trifce and Tomovski, Zivorad and Dubbeldam, Johan L A and Chechkin, Aleksei},
title = {Generalized diffusion-wave equation with memory kernel},
journal = {J. Phys. A: Math. Theor.},
}

@book{klages2008anomalous,
  title={Anomalous Transport: Foundations and Applications},
  author={Klages, R. and Radons, G. and Sokolov, I.M.},
  isbn={9783527622986},
  url={https://books.google.com/books?id=N1xD7ay06Z4C},
  year={2008},
  publisher={Wiley}
}

@article{METZLER20001,
title = {The random walk's guide to anomalous diffusion: a fractional dynamics approach},
journal = {Phys. Rep.},
volume = {339},
number = {1},
pages = {1--77},
year = {2000},
issn = {0370-1573},
doi = {10.1016/S0370-1573(00)00070-3},
author = {Ralf Metzler and Joseph Klafter}
}

@article{MainardiScalas2004,
  title={A fractional generalization of the Poisson processes},
  author={Francesco Mainardi and Rudolf Gorenflo and Enrico Scalas},
  journal={Vietnam J. Math.},
  volume={32},
  pages={53--64},
  year={2004},
  doi={10.48550/arXiv.math/0701454}
}

@article{Metzler_2004,
doi = {10.1088/0305-4470/37/31/R01},
year = {2004},
volume = {37},
number = {31},
pages = {R161},
author = {Ralf Metzler and Joseph Klafter},
title = {The restaurant at the end of the random walk: recent developments in the description of anomalous transport by fractional dynamics},
journal = {J. Phys. A Math. Gen.}
}

@article{HEOM_1,
author = {Tanimura ,Yoshitaka and Kubo ,Ryogo},
title = {Time Evolution of a Quantum System in Contact with a Nearly Gaussian-Markoffian Noise Bath},
journal = {J. Phys. Soc. Jpn.},
volume = {58},
number = {1},
pages = {101-114},
year = {1989},
doi = {10.1143/JPSJ.58.101},
}

@article{HEOM_2,
  title = {Nonperturbative expansion method for a quantum system coupled to a harmonic-oscillator bath},
  author = {Tanimura, Yoshitaka},
  journal = {Phys. Rev. A},
  volume = {41},
  issue = {12},
  pages = {6676--6687},
  numpages = {0},
  year = {1990},
  month = {Jun},
  publisher = {American Physical Society},
  doi = {10.1103/PhysRevA.41.6676},
  url = {https://link.aps.org/doi/10.1103/PhysRevA.41.6676}
}

@article{HEOM_3,
author = {Ishizaki ,Akihito and Tanimura ,Yoshitaka},
title = {Quantum Dynamics of System Strongly Coupled to Low-Temperature Colored Noise Bath: Reduced Hierarchy Equations Approach},
journal = {J. Phys. Soc. Jpn.},
volume = {74},
number = {12},
pages = {3131-3134},
year = {2005},
doi = {10.1143/JPSJ.74.3131},
}

@article{HEOM_4,
author = {Tanimura ,Yoshitaka},
title = {Stochastic Liouville, Langevin, Fokker–Planck, and Master Equation Approaches to Quantum Dissipative Systems},
journal = {J. Phys. Soc. Jpn.},
volume = {75},
number = {8},
pages = {082001},
year = {2006},
doi = {10.1143/JPSJ.75.082001},
}

@article{Dalibard1992,
  title = {Wave-function approach to dissipative processes in quantum optics},
  author = {Dalibard, Jean and Castin, Yvan and M\o{}lmer, Klaus},
  journal = {Phys. Rev. Lett.},
  volume = {68},
  issue = {5},
  pages = {580--583},
  numpages = {0},
  year = {1992},
  month = {Feb},
  publisher = {American Physical Society},
  doi = {10.1103/PhysRevLett.68.580}
}

@article{Dum1992,
  title = {Monte Carlo simulation of the atomic master equation for spontaneous emission},
  author = {Dum, R. and Zoller, P. and Ritsch, H.},
  journal = {Phys. Rev. A},
  volume = {45},
  issue = {7},
  pages = {4879--4887},
  numpages = {0},
  year = {1992},
  month = {Apr},
  publisher = {American Physical Society},
  doi = {10.1103/PhysRevA.45.4879}
}

@article{Plenio1998,
  title = {The quantum-jump approach to dissipative dynamics in quantum optics},
  author = {Plenio, M. B. and Knight, P. L.},
  journal = {Rev. Mod. Phys.},
  volume = {70},
  issue = {1},
  pages = {101--144},
  numpages = {0},
  year = {1998},
  month = {Jan},
  publisher = {American Physical Society},
  doi = {10.1103/RevModPhys.70.101}
}

@article{Breuer2007GeneralizedLindblad,
  title = {Non-Markovian generalization of the Lindblad theory of open quantum systems},
  author = {Breuer, Heinz-Peter},
  journal = {Phys. Rev. A},
  volume = {75},
  issue = {2},
  pages = {022103},
  numpages = {9},
  year = {2007},
  month = {Feb},
  publisher = {American Physical Society},
  doi = {10.1103/PhysRevA.75.022103},
  url = {https://link.aps.org/doi/10.1103/PhysRevA.75.022103}
}

@article{Makri1995QUAPI1,
  author = {Makri, Nancy and Makarov, Dmitrii E.},
  title = {Tensor propagator for iterative quantum time evolution of reduced density matrices. I. Theory},
  journal = {J. Chem. Phys.},
  volume = {102},
  pages = {4600--4610},
  year = {1995},
  doi = {10.1063/1.469508}
}

@article{Makri1995QUAPI2,
  author = {Makri, Nancy and Makarov, Dmitrii E.},
  title = {Tensor propagator for iterative quantum time evolution of reduced density matrices. II. Numerical methodology},
  journal = {J. Chem. Phys.},
  volume = {102},
  pages = {4611--4618},
  year = {1995},
  doi = {10.1063/1.469509}
}

@article{Thoss2001MCTDH,
  author = {Thoss, Michael and Wang, Haobin and Miller, William H.},
  title = {Self-consistent hybrid approach for complex systems: Application to the spin-boson model with Debye spectral density},
  journal = {J. Chem. Phys.},
  volume = {115},
  pages = {2991--3005},
  year = {2001},
  doi = {https://doi.org/10.1063/1.1385562}
}

@article{Wang2003MCTDHReview,
  author = {Wang, Haobin and Thoss, Michael},
  title = {Multilayer formulation of the multiconfiguration time-dependent Hartree theory},
  journal = {J. Chem. Phys.},
  volume = {119},
  pages = {1289--1299},
  year = {2003},
  doi = {10.1063/1.1580111}
}

\end{document}